%% file: main.tex
\documentclass[a4paper,usenatbib]{mnras}

\makeatletter

\newcommand{\Rmnum}[1]{\expandafter\@slowromancap\romannumeral #1@}
\newcommand{\lsim}{\lesssim}
\newcommand{\gsim}{\lower0.6ex\vbox{\hbox{$\buildrel{\textstyle >}\over{\sim}\ $}}}

\makeatother

\usepackage{newtxtext,newtxmath}
\usepackage{graphicx}	% Including figure files
\usepackage{amsmath}	% Advanced maths commands
\usepackage{amssymb}	% Extra maths symbols
\usepackage{subcaption}
\usepackage{color}

\title[Obscured quasars at $z \gtrsim 7$]{QSO obscuration at high redshift ($z \gtrsim 7$): Predictions from the BlueTides simulation}

\author[Y. Ni et al.]
{Yueying Ni$^{1}$\thanks{Email:yueyingn@andrew.cmu.edu}, Tiziana Di Matteo$^{1}$, Roberto Gilli$^{2}$, Rupert A.C. Croft$^{1}$, Yu Feng$^{3}$, \newauthor Colin Norman$^{4,5}$ \\
$^1$ McWilliams Center for Cosmology, Department of Physics, Carnegie Mellon University, Pittsburgh, PA 15213 \\
$^2$ INAF Osservatorio Astronomico di Bologna, Via Gobetti 93/3, I-40129 Bologna, Italy \\
$^3$ Berkeley Center for Cosmological Physics and Department of Physics, University of California, Berkeley, CA 94720, USA \\
$^4$ Space Telescope Science Institute, 3700 San Martin Dr., Baltimore, MD, 21218 USA \\
$^5$ Johns Hopkins University - Center for Astrophysical Sciences, 3400 N. Charles Street, Baltimore, MD 21218, USA \\
}

\date{Accepted XXX. Received YYY; in original form ZZZ}

\pubyear{2019}
\begin{document}
\maketitle

\begin{abstract}
High-$z$ AGNs hosted in gas rich galaxies are expected to grow through significantly obscured accretion phases. This may limit or bias their observability.
In this work, we use \textsc{BlueTides}, a large volume cosmological simulation of galaxy formation to examine quasar obscuration for the highest-redshift ($z \geq 7$) supermassive black holes residing in the center of galaxies. 
We find that for the bright quasars, most of the high column density gas ($>90\%$) resides in the innermost regions of the host galaxy, (typically within $< 10$ ckpc), while the gas in the outskirts is a minor contributor to the $N_\mathrm H$.
The brightest quasars can have large angular variations in galactic obscuration, over 2 orders of magnitude
(ranging from column density, $N_\mathrm H \sim 10^{21.5 - 24} \rm{cm}^{-2}$), where the lines of sight with the lowest obscuration are those formed via strong gas outflows driven by AGN feedback. 
We find that for the overall AGN population, the mean $N_\mathrm H$ is generally larger for high luminosity and BH mass, while the $N_\mathrm H$ distribution 
is significantly broadened, developing a low  $N_\mathrm H $ wing due to the angular variations driven by the AGN outflows/feedback. 
AGNs in high Eddington accretion phases are typically more heavily obscured.
We show that the linear relation between $\log N_{\rm H}$ and $\log L_X$ can be fit by $\log N_{\rm H} = (0.42 \pm 0.02) \log L_{X} + (5.1 \pm 0.7)$.
The obscured fraction P($N_{\rm H} > 10^{23} {\rm cm}^{-2}$) typically range from 0.6 to 1.0 for increasing $L_{X}$ (with $L_X > 10^{43} \rm{ergs/s}$), with no clear trend of redshift evolution.
With respect to the galaxy host property, we find that the $N_{\rm H}$ distribution peaks at higher value but also gets broadened and skewed toward low $N_{\rm H}$ for the hosts with larger stellar mass or molecular gas mass.
We find a linear relation between $N_{\rm H}$, $M_*$ and $M_{\rm H_2}$ with $\log N_{\rm H} = (0.24 \pm 0.03) \log M_{*} + (20.7 \pm 0.3)$ and $\log N_{\rm H} = (0.47 \pm 0.03) \log M_{\rm H_2} + (18.4 \pm 0.3)$. 
The dust optical depth in the UV band $\tau_{\mathrm UV}$ has tight positive correlation with $N_{\rm H}$. Our dust extincted UVLF is about 1.5 dex lower than the intrinsic UVLF, implying that more than 99\% of the $z \sim 7$ AGNs are heavily dust extincted and therefore would be missed by the UV band observation.
\end{abstract}


%%%%%%%%%%%%%%%%%%%%%%%%%%%%%%%%%%%%%%%%%%%%%%%%%%

%%%%%%%%%%%%%%%%% BODY OF PAPER %%%%%%%%%%%%%%%%%%

\input{Sec1_Introduction.tex}
\input{Sec2_Method.tex}
\input{Sec3_Result_P1.tex}

\input{Sec3_Result_P2.tex}

\input{Sec3_Result_P3.tex}

\input{Sec4_Conclusion.tex}

\section*{Acknowledgements}
We thank Fabio Vito for invaluable discussions and comments on this work.
YN thanks the help from Nick Gnedin for inspiring suggestions. 
The \textsc{BlueTides} simulation was run on the BlueWaters facility at the National Center for Supercomputing Applications.
TDM acknowledges funding from NSF ACI-1614853, NSF AST-1517593, NSF AST-1616168 and NASA ATP 19-ATP19-0084.
TDM and RACC also acknowledge funding from  NASA ATP 80NSSC18K101, and NASA ATP NNX17AK56G.

%%%%%%%%%%%%%%%%%%%%%%%%%%%%%%%%%%%%%%%%%%%%%%%%%%
%%%%%%%%%%%%%%%%%%%% REFERENCES %%%%%%%%%%%%%%%%%%
\bibliographystyle{mnras}
\bibliography{bib.bib}

%%%%%%%%%%%%%%%%%%%%%%%%%%%%%%%%%%%%%%%%%%%%%%%%%%
%%%%%%%%%%%%%%%%% APPENDICES %%%%%%%%%%%%%%%%%%%%%

\appendix
\section{Aitoff map of sample QSOs}
% \tiziana{You need to put some text and explanation here}
Here we give the full $N_\mathrm{H}$, $\tau_{\rm UV}$ and $v_r$ maps of three QSO samples BH2, BH3
and BH4 (repeating the format of Figure~\ref{fig:BH0map}) as an extension of Figure~\ref{fig:Nhmaps}, 
to further illustrate the correlation between the direction with low column density and high outward velocity. 

\label{sec:Appendix}

\begin{figure*}
    \begin{subfigure}[b]{0.48\textwidth}
        \includegraphics[width=\linewidth]{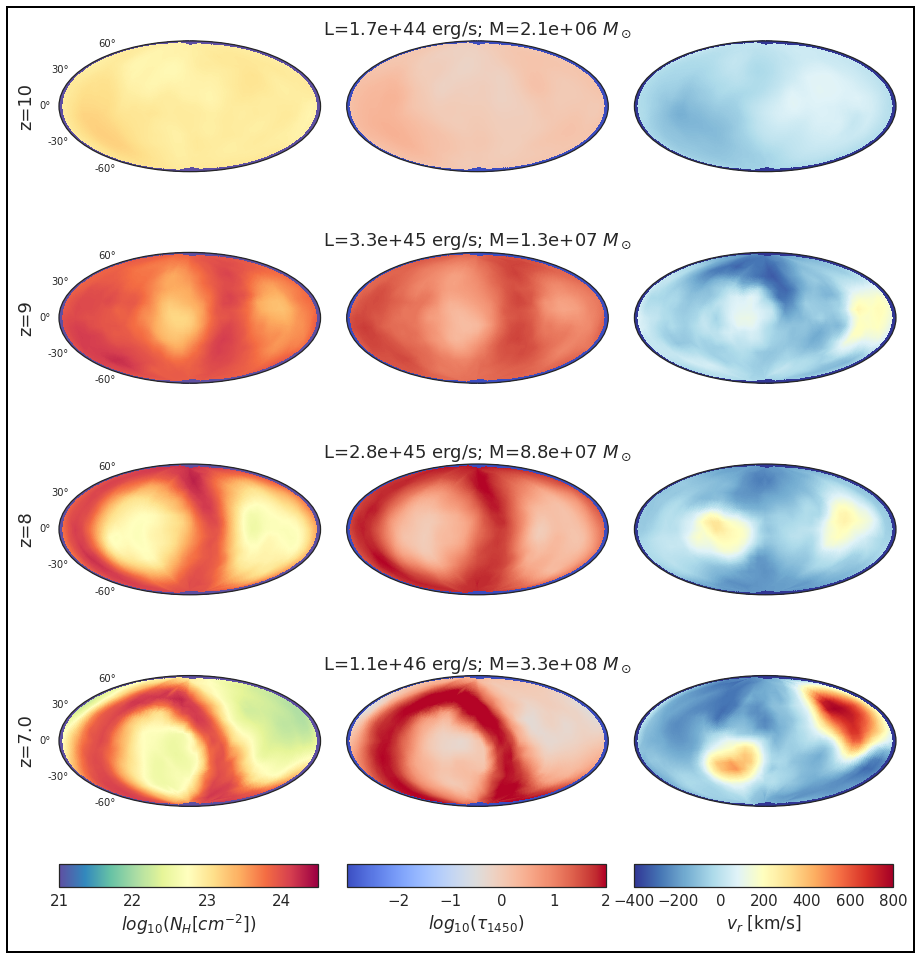}
        \caption{BH2}
    \end{subfigure}
    ~
    \begin{subfigure}[b]{0.48\textwidth}
        \includegraphics[width=\linewidth]{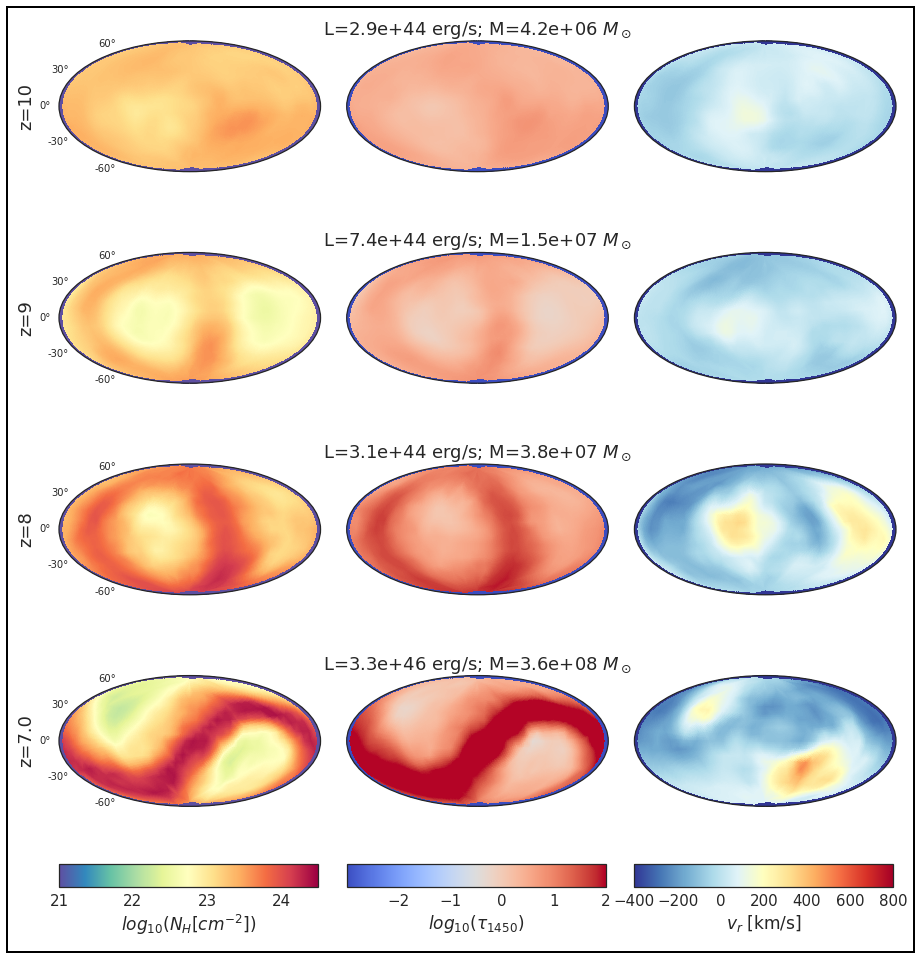}
        \caption{BH3}
    \end{subfigure}
    
    \begin{subfigure}[b]{0.48\textwidth}
        \includegraphics[width=\linewidth]{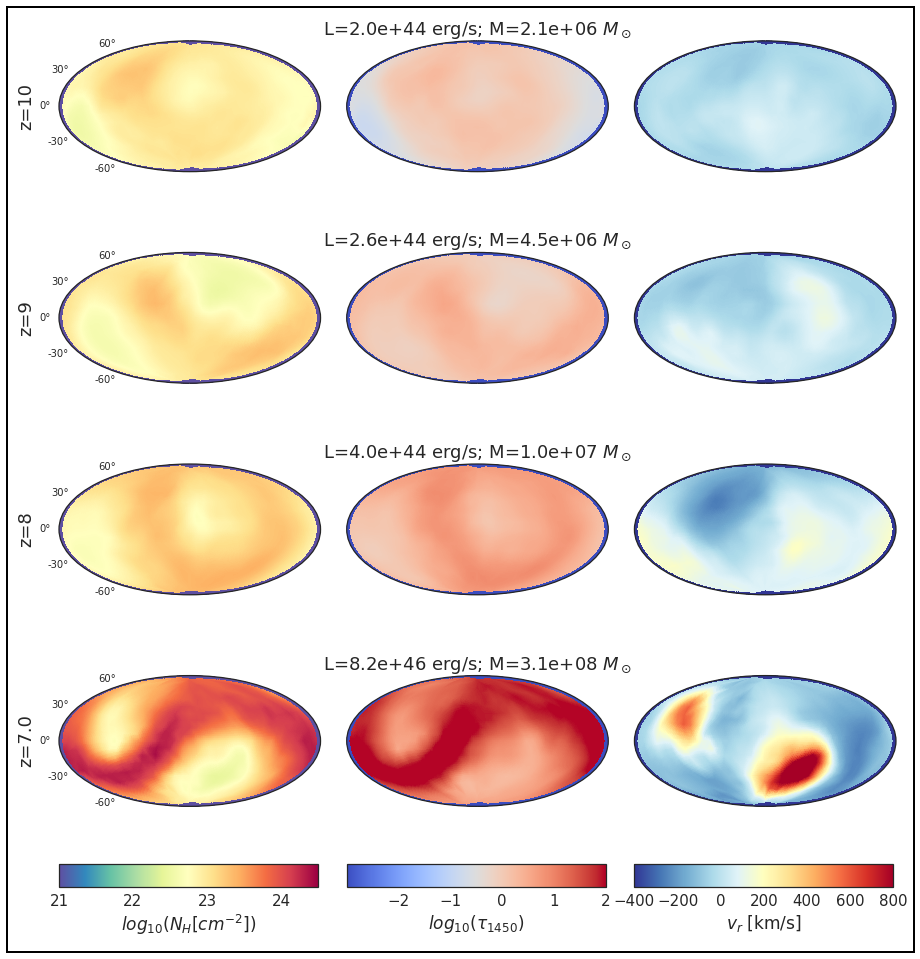}
        \caption{BH4}
    \end{subfigure}
    
    \caption{
    Same as Figure~\ref{fig:BH0map}, here we plot the map of $N_\mathrm{H}$, $\tau_{\rm UV}$ and averaged radial velocity $v_r$ of the other 3 sample QSOs. 
    Each row represent the state at a certain redshift, evolving from $z=10$ to $z=7$ (from top to bottom). 
    The left-most column is the map of hydrogen column density $N_\mathrm{H}$. The middle column is the map of dust optical depth $\tau_{\rm UV}$. The right column gives the averaged radial velocity $v_r$ in each line of sight. The title gives the BH bolometric luminosity and BH mass at the corresponding redshift.
    }
\label{fig:Nhmaps-3others}
\end{figure*}

%%%%%%%%%%%%%%%%%%%%%%%%%%%%%%%%%%%%%%%%%%%%%%%%%%
% Don't change these lines
\bsp	% typesetting comment
\label{lastpage}

\end{document}

%% file: Sec1_Introduction.tex
\section{Introduction}

\label{section1:introduction}

Understanding the origin of the first supermassive black holes (SMBHs) powering the most luminous quasars (QSOs) at $z>6$ is one of the greatest observational and theoretical challenges. 
To understand the formation and growth of these SMBHs as well as their co-evolution with their host galaxy, one needs to consider that while they are extremely massive for this early epoch, their space density is extremely low.
The inferred rarity of these high-$z$ QSOs may however be partly due to large amounts of star forming gas in these early host galaxies that may act to obscure an active galactic nucleus (AGN).

The obscuring medium for AGN is typically composed of dust and/or gas. 
Dust is the dominant source of obscuration in UV-IR bands while gas dominates the absorption at X-ray energies \citep{Hickox2018}.
% The typical definition of obscured AGN corresponds to a typical extinction from dust with $A_v = 5- 10$ mag. 
For moderate column densities, detection of the X-ray emission is an efficient way to reveal the presence of an AGN (that is obscured in the UV). However, when the equivalent neutral hydrogen column exceeds the unit optical depth corresponding to the Thompson cross section ($N_{\rm H} > 1.5 \times 10^{24} \mathrm{cm}^{-2}$), such "Compton-thick" AGN  are hard to detect even in X-ray surveys.
% due to Compton recoil and subsequent absorption of X-ray photons \citep{Comastri2004}.

The obscuration of AGN can occur over a range of scales and physical conditions. 
In the standard unification scheme \citep[e.g.][]{Antonucci1993,Urry1995}, AGNs are surrounded by an optically thick toroidal structure a few parsec from the central BH. This torus then obscures the line of sight depending on the viewing angle.
Alternatively, or additionally, obscuration can be produced by gas on the scale of the entire galaxy (>kpc) \citep{Maiolino1995}.
% Higher redshift AGNs residing in star-forming galaxies are often rich in gas, ...
High redshift AGNs ($z$ = 1-3) residing in star-forming galaxies are often rich in gas \citep[see, e.g.][]{Tacconi2013}, 
and the obscuration is thought to be caused by the high density gas fueling the star formation as well as the accretion process.
Recently, X-ray observations of $z>2.5$ obscured AGN \citep{Circosta2019} indeed confirm this picture, 
whereby at high redshift, the host ISM (at kpc scale) is dense enough to produce the inferred amount of X-ray absorption, explaining
also the increased fraction of obscured AGN  at high-$z$.
% significant absorption is dominated by the dense ISM in the host galaxy (at kpc scale).
% adds to or substitutes that produced by the torus in circumnuclear region (at pc scale).

% Roberto
% what we could really say in Circosta+19 is that the host ISM density is enough to produce the observed X-ray absorption. While we believe this may be the case, unfortunately is not a direct demonstration that it does.

The recent observations of the highest redshift, $z>7$, quasars imply BH masses of order $10^9 M_{\odot}$. In order to grow to such high mass in the first few $10^8$ yrs of the universe, those SMBHs need to undergo continuous near-Eddington or even super-Eddington accretion, during which they are enshrouded by accreting gas with a column density that exceeds even the Compton-thick level \citep[e.g.][]{Pezzulli2017}.
Observations of Ly$\alpha$ absorption profiles from two $z>7$ QSOs \citep{Davies2019} imply that they might have experienced highly obscured growth and could be obscured in more than 82\% of their lifetimes (with assumption of similar radiative efficiency as the low redshift QSOs.)

Different studies have been carried out to assess the fraction of obscured AGNs (including Compton-thick AGNs) as a function of AGN luminosity and redshift \citep[see,][for a review]{Hickox2018}.
The correlation between obscured fraction and luminosity is still under debate, however.
Some studies of lower redshift ($z<3$) quasars indicate that the obscured AGN fraction decreases toward higher luminosity \citep[e.g.][]{Merloni2014} or higher Eddington ratio \citep{Ricci2017},
while there are also works that claim that the luminosity dependence is not quite significant \citep[e.g.][]{Mateos2017}. 
There is also evolution of obscured AGN fraction with redshift: in particular, the obscured fraction of $3<z<6$ luminous AGN is found to increase when going to higher redshift \citep{Vito2014,Vito2018}.
% The redshift evolution appears to be stronger at high luminosity,
% which is the reason why we see a flat obscured fraction at all luminosities at z>3.
% For more luminous AGN the obscured fraction to increase with redshift 
% \tiziana{please quote numbers and z!}

The studies referenced above of the obscured AGN fraction are mostly based on the AGN population at $z \lsim 5$.
There is rapid ongoing progress in the detection of $z>6$ and even $z>7$ QSOs from observations:
more than 200 QSOs have been discovered beyond $z=6$ \citep[see,e.g.][and references therein]{Fan2019}, and a handful found with $z>7$ \citep{Mortlock2011,Wang2018,Banados2018,Matsuoka2019,Yang2019}.
Although such studies represent a huge breakthrough, the current detections of these $z>7$ quasars are mostly obtained from wide-field optical/near-IR surveys, which as a  selection method is strongly biased against highly obscured systems \citep[e.g., see discussion in][]{Vito2019}.
Studies of AGN populations at lower redshift ($0<z<5$) show that the vast majority of the AGN population is obscured \citep{Ueda2014,Buchner2015}. 
Recently, the first heavily obscured $z=6.5$ AGN candidate has been discovered through \textit{Chandra} X-ray survey \citep{Vito2019}.
The hardness of the X-ray photometry leads to an inferred galactic absorption of up to $N_{\mathrm H} > 2 \times 10^{24} \mathrm{cm}^{-2}$ and $N_{\mathrm H} > 6 \times 10^{23} \mathrm{cm}^{-2}$ at the $68\%$ and $90\%$ confidence levels respectively.

Making theoretical predictions for the obscured fraction of the quasar population is essential to reach a complete census of high-$z$ QSOs and further understand the formation and growth of SMBH in the early Universe.
In this paper, we use the \textsc{BlueTides} simulation to study the obscuration state of $z>7$ QSOs in
a $\Lambda$CDM universe.
\textsc{BlueTides} is a cosmological hydrodynamic simulation targeted at the study of the first generation of galaxies and QSOs in the high-$z$ universe \citep{DiMatteo2017,Tenneti2018,Ni2018}.
The large volume and high resolution of \textsc{BlueTides}, 
% which is almost 200 times larger than either the Illustris~\citep{Vogelsberger2014}, or EAGLE~\citep{Schaye} simulations,
makes it ideally suited to study the rare high-$z$ luminous QSOs, along with the detailed structure and physical properties of their surrounding gas.
So far, \textsc{BlueTides} has been tested against various observations  of the high-$z$ universe and has been shown to be in good agreement with all current observational constraints, such as the UV luminosity functions \citep{Feng2016, Waters2016a, Waters2016b, Wilkins2017}, the first galaxies and the most massive quasars \citep{Feng2015, DiMatteo2017, Tenneti2018}, the Lyman continuum photon production efficiency \citep{Wilkins2016, Wilkins2017}, galaxy stellar mass functions \citep{Wilkins2018}, angular clustering amplitude \citep{Bhowmick2017}, BH-galaxy scaling relations \citep{Huang2018}, and gas outflows from the $z=7.54$ quasar \citep{Ni2018}.

In this work, we focus on the galactic obscuration that is due to ISM gas in the QSO host galaxies. 
%Therefore, our estimation actually sets a lower limit to the total obscuration.
Numerical zoom-in simulations can better resolve the obscuring gas in the nuclear region of SMBH \citep[see e.g.][]{Hopkins2016,Trebitsch2019, Lupi2019}.
% In separate studies, using numerical zoom-in simulations
% the obscuration in the nearby environment of a SMBH has been studied \citep[e.g.][]{Hopkins2016,Trebitsch2019} for particular halos/host galaxies.  
In particular, \cite{Trebitsch2019} has found that the gas from the host galaxy contributes to the total obscuration at a level at least comparable to the gas in the nuclear region.
Given the significant role played by galactic obscuration for high-$z$ QSOs, the tens of thousands of SMBHs contained in the \textsc{BlueTides} simulation at $z>7$ enable us to study the statistics of galactic obscuration and the emerging BH population at high redshift. We are particularly interested in how current detections of high-z quasars may be hampered by obscuration.

This paper is organized as follows. 
In Section~\ref{section2:Method}, we briefly summarize the sub-grid models applied in \textsc{BlueTides} and describe the model we use to calculate the quasar and galaxy luminosity. 
We also introduce the way we calculate the gas and dust obscuration through AGN lines of sight.
Section~\ref{section3:result} show and discuss our key results.
In Section~\ref{section4:Summary}, we conclude the paper.

%% file: Sec2_Method.tex
\section{Method}
\label{section2:Method}

% Subsection 2.1  ----------------------------------------------------------------------------------------------------
\subsection{BlueTides simulation}

The \textsc{BlueTides} \footnote{http://BlueTides-project.org/} cosmological simulation~\citep{Feng2016} uses the Pressure Entropy Smoothed Particle Hydrodynamics code MP-Gadget to model the evolution of a $400 \rm{Mpc}/h$ side box with $2\times 7040^{3}$ particles.
The simulation evolved from $z=99$ and has now reached $z < 7$. 
The cosmological parameters used are from the nine-year Wilkinson Microwave Anisotropy Probe (WMAP) \citep{Hinshaw2013} ($\Omega_0=0.2814$, $\Omega_\Lambda=0.7186$, $\Omega_b=0.0464$, $\sigma_8=0.82$, $h=0.697$, $n_s=0.971$).

\textsc{BlueTides} implements a variety of sub-grid models to model galaxy formation and different feedback process.
Here we briefly list some of its basic features, and we refer the reader to the original papers~\citep{Feng2016} for detailed descriptions. 
In the simulations, gas is allowed to cool through both radiative processes~\citep{Katz} and metal cooling. 
The metal cooling rate is obtained by scaling a solar metallicity template according to the metallicity of gas particles, following the method described in \cite{Vogelsberger2014}.
Star formation (SF) is based on a multi-phase SF model ~\citep{SH03} with modifications following~\cite{Vogelsberger2013}.
We model the formation of molecular hydrogen and its effects on SF at low metallicity according to the prescription of \cite{Krumholtz}. 
We self-consistently estimate the fraction of molecular hydrogen gas from the baryon column density, which in turn couples the density gradient to the SF rate.
Type II supernova wind feedback (the model used in Illustris ~\citep{Nelson}) is included, assuming wind speeds proportional to the local one dimensional dark matter velocity dispersion. 
The large volume of \textsc{BlueTides} also allows to include a model of "patchy reionization" ~\citep{Battaglia}, yielding a mean reionization redshift $z\sim10$, and incorporating the UV background estimated by \cite{fg09}. 

In our simulation, we model BH growth and AGN feedback in the same way as in the \textit{MassiveBlack} $I \& II$ simulations, using the BH sub-grid model developed in \cite{SDH2005,DSH2005} with modifications consistent with \textit{Illustris}. 
BHs are seeded with an initial seed mass of $M_{\mathrm {seed}} = 5 \times 10^5 h^{-1} M_{\odot}$ (commensurate with the resolution of the simulation) in halos with mass more than $5 \times 10^{10} h^{-1} M_{\odot}$. 
The gas accretion rate onto BH is given by a modified Bondi accretion rate,
\begin{equation}
    \dot{M}_B = \frac{4 \pi \alpha G^2 M_{\rm BH}^2 \rho}{(c^2_s+v_{\rm rel}^2)^{3/2}}
\end{equation}
where $c_s$ and $\rho$ are the local sound speed and density of gas, and $v_{\rm rel}$ is the relative velocity of the black hole to the nearby gas. 
We allow for super-Eddington accretion in the simulation but limit the accretion rate to 2 times the Eddington accretion rate:
\begin{equation}
\label{equation:Meddington}
    \dot{M}_{\rm Edd} = \frac{4 \pi G M_{\rm BH} m_p}{\eta \sigma_{T} c}
\end{equation}
where $m_p$ is the proton mass, $\sigma_T$ the Thompson cross section, c is the speed of light, and $\eta=0.1$ is the radiative efficiency of the accretion flow onto the BH.
Therefore, the BH accretion rate in \textsc{BlueTides} simulation is determined by:
\begin{equation}
    \dot{M}_{\rm BH} = {\rm Min} (\dot{M}_B, 2\dot{M}_{\rm Edd})
\end{equation}
The Eddington ratio defined as $\lambda_{\mathrm {Edd}}= \dot{M}_{\mathrm {BH}}$/2$\dot{M}_{\mathrm {Edd}}$ would usually range from $0 \sim 1$ during the evolution of AGN.

The SMBH is assumed to radiate with a bolometric luminosity $L_{\rm Bol}$ proportional to the accretion rate $\dot{M}_{\rm BH}$:
\begin{equation}
    L_{\rm Bol} = \eta \dot{M}_{\rm BH} c^2
\end{equation}
with $\eta = 0.1$ being the mass-to-light conversion efficiency in accretion disk according to \cite{Shakura1973}.
5\% of the radiation energy is thermally coupled to the surrounding gas that resides within twice the radius of the SPH smoothing kernel of the BH particle. This
scale is typically about 1\% $\sim$ 3\% of the virial radius of the halo.
The AGN feedback energy only appears in kinetic form through the action of
this thermal energy deposition, and no other coupling (e.g.,radiation pressure) is included.

% Subsection 2.1  ----------------------------------------------------------------------------------------------------

% Fig ----------------------------------------------------------------------------------------------
\begin{figure*}
\centering
\includegraphics[width=2.0\columnwidth]{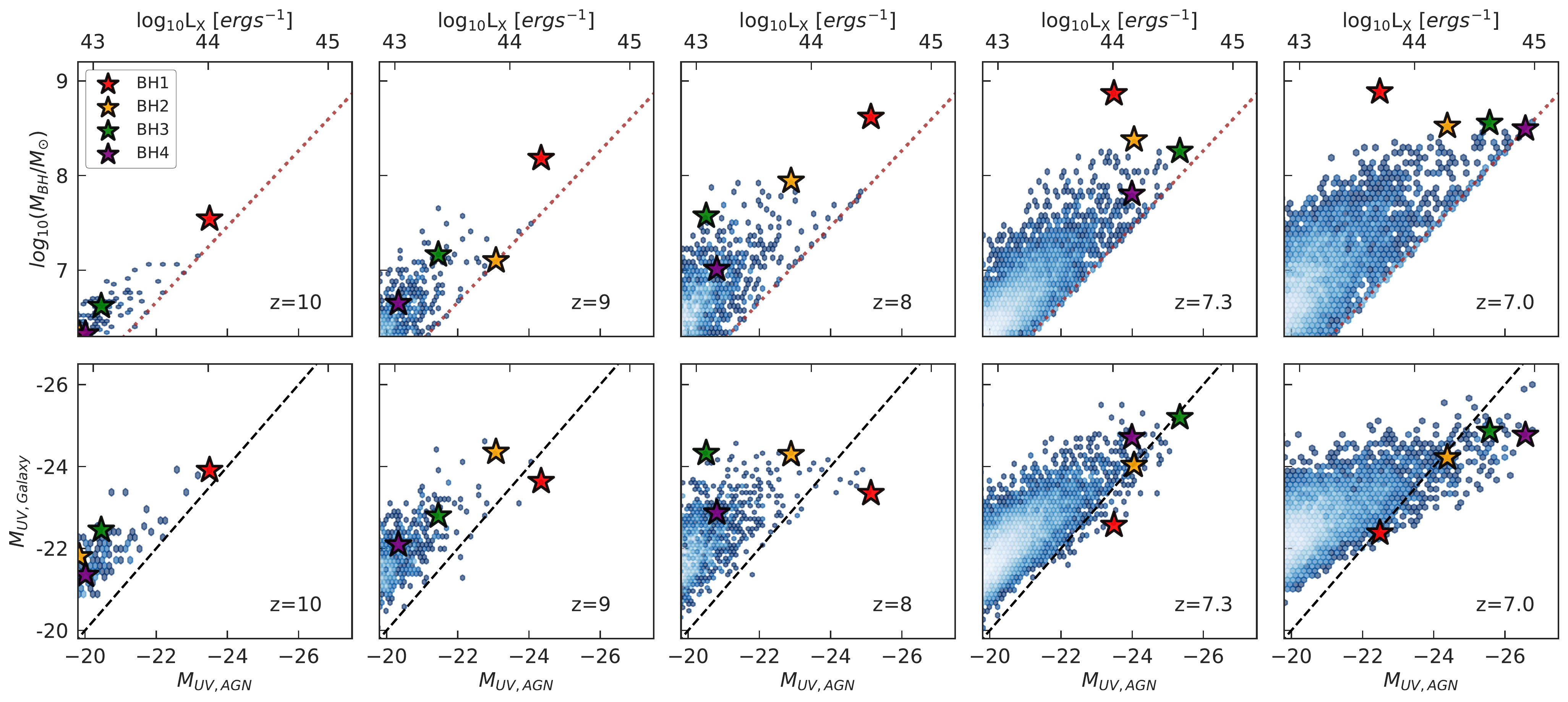}
\caption{
Properties of the SMBH population and their host galaxies in the \textsc{BlueTides} simulation measured from $z=10$ to $z=7.0$ (from left to right).
The $x$ axis shows the QSO luminosity. The upper $x$ axis shows the corresponding (intrinsic) quasar luminosity in X-ray band, and the lower $x$ axis shows the corresponding luminosity in UV band.
The colored stars mark the 4 sample QSOs we have selected for illustration in other
figures.
\textit{Top panel}:  2D histogram of BH mass and BH luminosity. The brown dotted lines are $2 \times L_{\mathrm{Edd}}(M_{\mathrm{BH}})$ for the corresponding BH mass, which represents  the upper bound luminosity for a certain BH mass set in the simulation.
\textit{Bottom panel}:
2D histograms of the UV band (intrinsic) luminosity of AGN and their host galaxies. Black dashed lines indicate where the AGN and the host galaxy have the same UV band luminosity.
}
\label{fig:Muv_collection}
\end{figure*}

\begin{figure*}
% \vspace{-1cm}
\includegraphics[width=2.1\columnwidth]{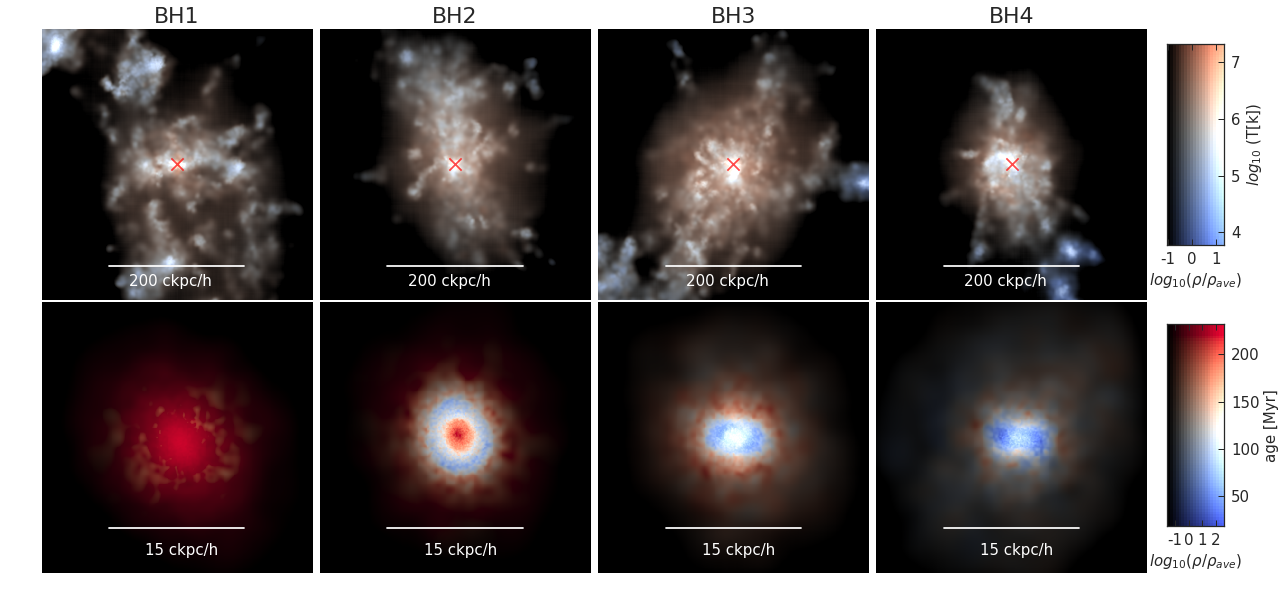}
    \caption{
    Illustration of the host environment of the four sample QSOs at $z=7.0$. 
    The top panels give the gas density field color coded by temperature (blue to red indicating cold to hot respectively, as shown by the color bar aside).
    % \rupert{Give degrees K values for the hot and cold}). 
    The boxes are 400 ckpc/h (comoving unit) per side, with BHs residing in the center (marked by the red cross in the top panels).
    Bottom panels paint the stellar density field for the host galaxy color coded by the age of stars (from blue to red indicating young to old populations respectively). They are 30 ckpc/h per side.}
    \label{fig:images}
\end{figure*}

% Table --------------------------------------------------------------------------------------------
\begin{table}
\label{tab:examples}
    \centering
    \begin{tabular}{lcccc}
    	\hline
    	  & BH1 & BH2 & BH3 & BH4 \\
    	\hline
    	$L_{\mathrm{Bol,BH}}$ [erg/s] & $1.9 \times 10^{45}$ & $1.1 \times 10^{46}$ & $3.3 \times 10^{46}$ & $8.2 \times 10^{46}$ \\
    	$M_{\mathrm{BH}}$ $[M_{\odot}]$ & $7.7 \times 10^{8}$ & $3.3 \times 10^{8}$ & $3.6 \times 10^{8}$ & $3.1 \times 10^{8}$ \\
    	\hline
    	$M_{*}$ $[M_{\odot}]$ & $4.2 \times 10^{10}$ & $6.8 \times 10^{10}$ & $8.7 \times 10^{10}$ & $4.8 \times 10^{10}$ \\
    	$M_{\mathrm{gas}}$ $[M_{\odot}]$ & $3.8 \times 10^{10}$ & $5.1 \times 10^{10}$ & $8.3 \times 10^{10}$ & $6.5 \times 10^{10}$ \\
    	$M_{\mathrm{H_2}}$ $[M_{\odot}]$ & $6.2 \times 10^{9}$ & $1.7 \times 10^{10}$ & $2.9 \times 10^{10}$ & $2.7 \times 10^{10}$ \\
    	\hline
    	$M_{200}$ $[M_{\odot}]$ & $6.3 \times 10^{11}$ & $6.5 \times 10^{11}$ & $9.0 \times 10^{11}$  & $6.2 \times 10^{11}$ \\
    	$R_{200}$ [kpc] & 22.2 & 22.4 & 25.0 & 22.1 \\
    	\hline
    \end{tabular}
    \caption{The host properties of the 4 example QSOs at $z=7$. Here $M_*$, $M_{\rm gas}$, and $M_{\rm H_2}$ are all calculated within the virial radius $R_{200}$ of the AGN host. 
    Here $R_{200}$ is given in physical units.
    }
\end{table}
% Table --------------------------------------------------------------------------------------------

% Subsection 2.2  ----------------------------------------------------------------------------------------------------
\subsection{Luminosity of AGN and galaxies}
\label{subsection:subsection 2.2}

To calculate the UV band luminosity of AGN, we apply bolometric corrections to convert the $L_{\rm Bol}$ to rest frame UV band absolute magnitude $M_{\rm UV}$ following \cite{Fontanot2012}
\begin{equation}
    M_{\rm UV} = -2.5 \rm{log_{10}} \frac{L_{\rm Bol}}{f_B \mu_B}+34.1+\Delta_{\rm {B,UV}}
\end{equation}
where $f_{B}=10.2$, $\mu_B = 6.7 \times 10^{14}$Hz and $\Delta_{\mathrm{B,UV}}$=-0.48.

The UV luminosity of the host galaxy is obtained by modelling its spectral energy distribution, which is constructed by attaching the SED of a simple stellar population (SSP) to each star particle based on the age and metallicity. 
We employ version 2.1 of the Binary Population and Spectral Populations Synthesis (SPS) model \citep{Eldridge2017} utilizing a modified Salpeter IMF (Salpeter high-mass slope with a break at < $0.5 M_{\odot}$) and a high-mass cut-off of $100 M_{\odot}$.

We also convert AGN $L_{\rm Bol}$ to the luminosity in the hard X-ray band [2-10]~keV following the bolometric correction $L_X = L_{\rm Bol}/k$ from \citep{Hopkins2007}, with
\begin{equation}
    k(L_{\rm Bol}) = 10.83(\frac{L_{\rm Bol}}{10^{10}L_\odot})^{0.28}+6.08(\frac{L_{\rm Bol}}{10^{10}L_\odot})^{-0.020}
\end{equation}
here $L_{\odot}$ refers to the bolometric solar luminosity $L_{\odot}=3.9 \times 10^{33}$~ergs/s.

% Subsection 2.2  ----------------------------------------------------------------------------------------------------

\subsection{Gas obscuration around AGN}

Bright quasars in a high accretion state are typically shrouded by large fraction of high density gas, and the luminosities of quasars are largely reduced due to scatter and absorption by gas between them and an observer.
The gas density field around AGN (and the corresponding obscuration) can be a complex environment with large spatial and time variations due to  accretion and AGN feedback process.
In this work we explore in detail the angular variations of the gaseous environment of simulated  AGN and study the resulting obscuration on a statistical basis.

For each AGN, we use \textsc{Healpy} \footnote{https://healpy.readthedocs.io} to cast 972 evenly distributed lines of sight starting from the position of each AGN and calculate the hydrogen column density $N_{\rm H}$, UV optical depth $\tau_{\rm UV}$ and density averaged radial velocity $v_r$ for each line of sight.
More specifically, first we determine the gas density field using the SPH formalism 
\begin{equation}
\label{equation:rho_i}
    \rho(\mathbf{r}_i) = \sum_{j} m_j W_{ij} =  \sum_{j} m_j W(|\mathbf{r}_i - \mathbf{r}_j|,h_j),
\end{equation}
where $\sum_j$ is a sum over all the neighbouring gas particles within the smoothing length $h$, and $W$ is the quintic kernel used in \textsc{BlueTides}.
We then calculate $N_{\rm H}$ by integrating the hydrogen number density $n_{\rm H}$ along each line of sight: 
\begin{equation}
\label{equation:NH}
    N_{\rm H} =  \int_{\rm ray} X\rho(l)/m_p dl
\end{equation}
with $m_p$ the proton mass and $X = 0.76$ the hydrogen mass fraction.

As is usual in most cosmological simulations, \textsc{BlueTides} applies a uniform UV radiation background to all gas particles. It therefore  does
not include a self-consistent calculation of the neutral hydrogen fraction for gas in the environment of quasars. In order to account 
for the effect of the local quasar ionizing radiation, we therefore do the following: First, because only neutral hydrogen should contribute to $N_{\rm H}$, it is a good assumption that the star-forming gas is self-shielded from the photon-dissociating UV background and remains neutral. 
Therefore we only include  gas particles with non-zero star formation rate when calculating $N_{\rm H}$.  Second, we also assume that any gas which is 
not dense enough be forming stars is fully ionized by the nearby quasar.
We note that star forming gas resides in high density regions and dominates the contribution of $N_{\rm H}$, and we find as expected that there is negligible difference in our results if instead we compute $N_{\rm H}$ from all gas components.

We also quantify the averaged radial velocity for each line of sight and study its relationship with $N_{\rm H}$. 
To do this, we determine the velocity field using $\mathbf{v}_i = \sum_{j} m_j \mathbf{v}_j W_{ij}/ \rho_i$ and then calculate the averaged radial velocity $v_r$ weighted by density for each line of sight. 
 $v_r$ therefore acts as the momentum flux along the specified direction. We use this quantity as a proxy for gas outflow rate in later sections.

To study the AGN obscuration in the UV band brought about by dust attenuation, we employ a scheme same as \cite{Wilkins2017}, assuming that the metal density integrated along the line of sight is proportional to dust optical depth:
\begin{equation}
\label{equation:tau_dust}
    \tau_{\rm UV, AGN} = \kappa \left(\frac{\lambda}{5500\text{\normalfont\AA}}\right)^{\gamma} \int_{\rm ray} \rho_{\rm metal}(l) dl ,
\end{equation}
where $\rho_{\rm{metal}, i} = \sum_{j} m_j Z_j W_{ij}$ and $Z_j$ is the metallicity (in unit of mass fraction) of gas particle $j$. 
Here $\kappa = 10^{4.6}$ and $\gamma=-1.0$ is a free parameter that is calibrated against the observed galaxy luminosity function \citep[see also][]{Marshall2019}.
This method is well established in previous studies on luminous galaxy population of \textsc{BlueTides} simulation \citep[see, e.g.][for more detailed descriptions.]{Wilkins2017}
% \tiziana{this will be 'submitted'. Also 
% we can refer to Wilkins papers (where we already have already used this model to constrain $\kappa$ at $z=8$. This would just support the validity of the method}.  

% We can also quantify the dust attenuation in terms of the optical depth $\tau_d (\lambda)$, with the assumption that it is proportional to the metal column density:
% \begin{equation}
% \label{equation:tau_dust}
%     \tau_d(\lambda) = \int_{\rm ray} n_d(l) \sigma_d(\lambda) dl
% \end{equation}
% where $n_d$ is the number density of dust grains and $\sigma_d$ is the cross section of dust within certain wavelength.
% To calculate the dust absorption cross section $\sigma_d$ at UV band,
% we apply the dust model for Small Magellanic Clouds (SMC) following the work of \cite{Gnedin2008}, end up with $\sigma_d(1450\angstrom)=3.26 \times 10^{-22} \mathrm{cm}^2$ per hydrogen atom at the SMC metallicity.
% Then the dust number density $n_d$ is estimated via $n_d = n_{\rm H} Z/Z_0$, where $Z_0 = 0.005$ is the mean gas-phase metallicity of the SMC.

%% file: Sec3_Result_P1.tex
\section{Results}
\label{section3:result}
% Fig --------------------------------------------------------
% --------------------------------------
\begin{figure*}
    \begin{subfigure}[b]{0.48\textwidth}
        \includegraphics[width=\linewidth]{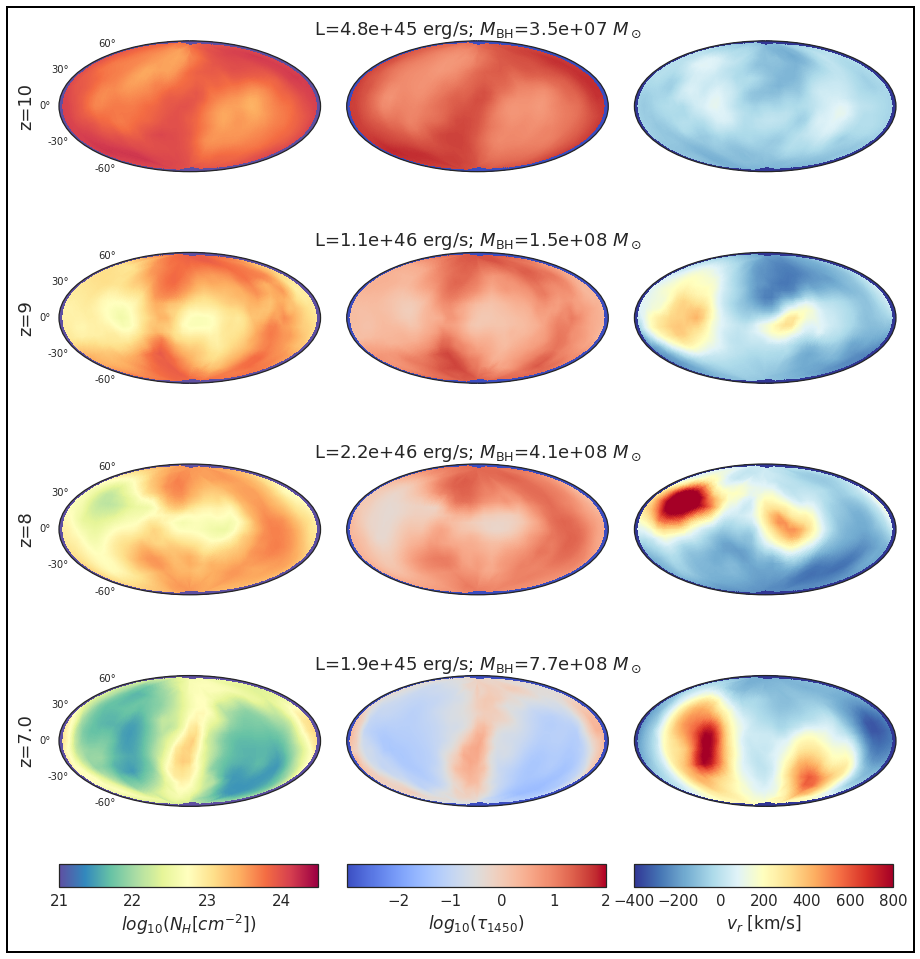}
        \caption{Maps of $N_{\rm H}$, dust optical depth $\tau_{1450}$ and radial velocity $v_r$ around BH1.}
        \label{fig:BH0map}
    \end{subfigure}
    ~
    \begin{subfigure}[b]{0.48\textwidth}
        \includegraphics[width=\linewidth]{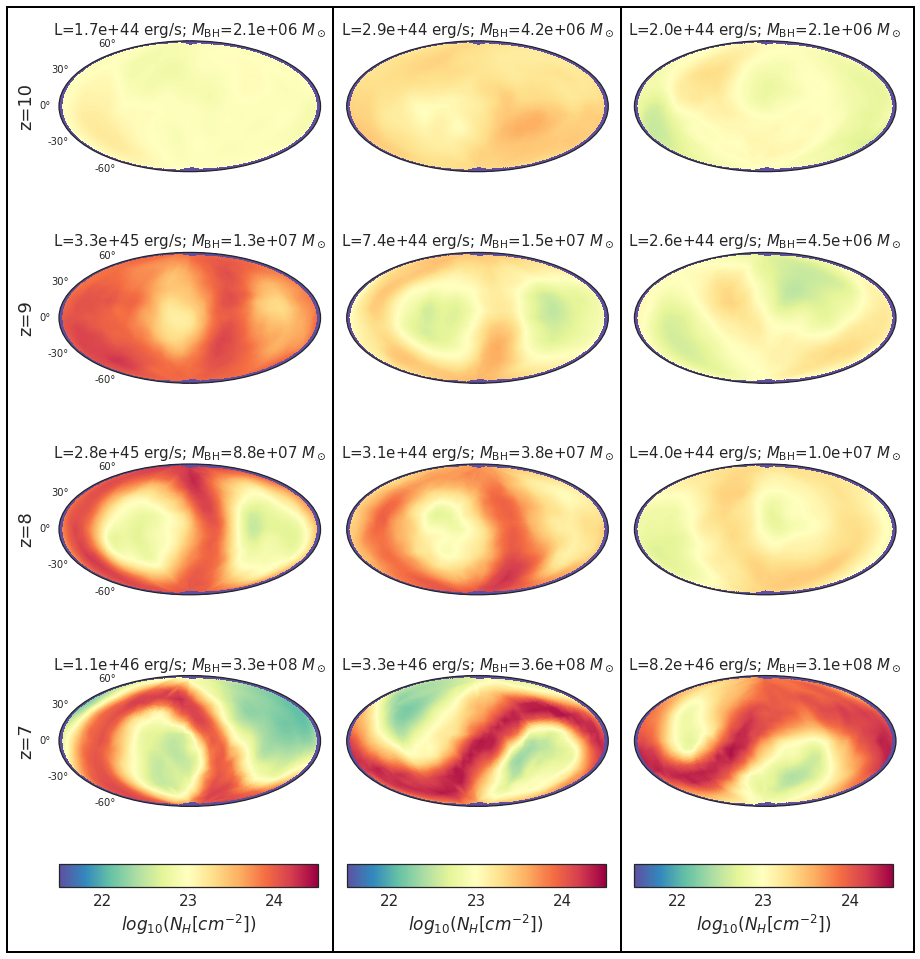}
        \caption{Maps of  $N_{\rm H}$ around black holes BH2, BH3, and BH4}
         \label{fig:3Nhmap}
    \end{subfigure}
    
    \caption{
    \textit{Left figurel}: 
    The properties of gas surrounding BH1 (the most massive BH) in an Aitoff projection. Each row represents a specific property at a certain redshift, and the columns its associated evolution from $z=10$ to $z=7$ (from top to bottom). 
    The left-most column is the map of hydrogen column density $N_\mathrm{H}$. The middle column is the map of dust optical depth $\tau_{\rm UV}$. The right column gives the averaged radial velocity $v_r$ along each line of sight. The associated  BH bolometric luminosity and BH mass at the corresponding redshift is given at the top of each panel.
    ~
    \textit{Right figure}: Maps showing the evolution of $N_\mathrm{H}$ with redshift for BH2 (left), BH3 (middle), BH4 (right).
    The full $N_\mathrm{H}$, $\tau_{\rm UV}$ and $v_r$ maps of these three sample QSOs are given in Appendix \ref{sec:Appendix}.
    Each panel is labeled by the BH bolometric luminosity and BH mass at the corresponding redshift.
    The maps at $z=7$ exhibit a bimodal morphology with regard to the angular distribution of the $N_\mathrm{H}$ field.
    }
\label{fig:Nhmaps}
\end{figure*}

\begin{figure*}
\includegraphics[width=2.1\columnwidth]{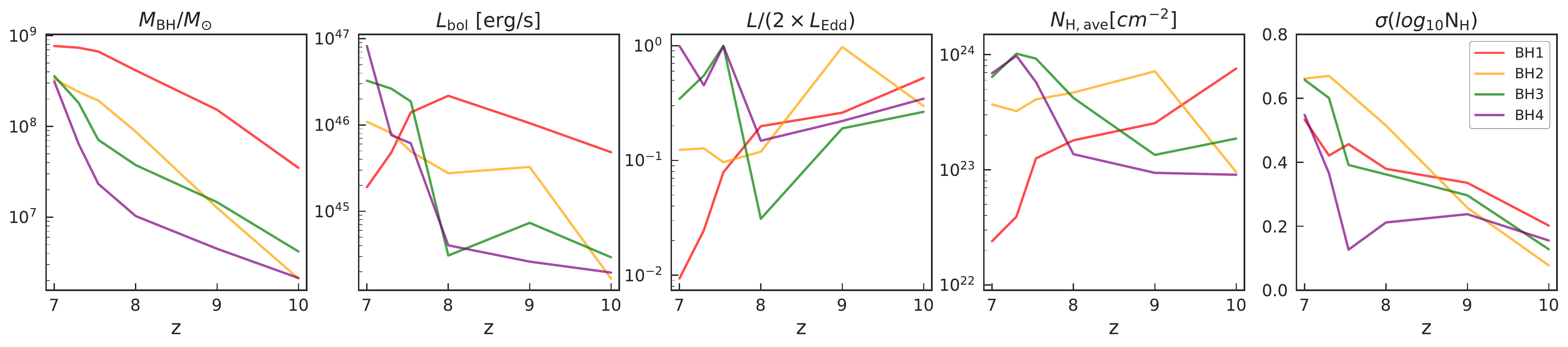}
    \caption{
    The time evolution of the four samples from $z=10$ to $z=7$.
    The first three panels show the following properties of each BH as a function of redshift: BH mass, bolometric luminosity of AGN, and the Eddington accretion ratio.
    The rightmost two panels quantify the obscuration state of the host environment.
    The fourth panel shows $N_{\rm H,ave}$, the $N_{\rm H}$ value averaged over all lines of sight.
    % Middle panel gives the probability of getting lines of sight with $N_{\rm H}>10^{23} \rm{cm}^{-2}$.
    The rightmost panel displays $\sigma(\rm log_{10} N_{\rm H})$ calculated from all lines of sight, as a simple measure of the angular variation in $N_{\rm H}$ values. 
    }
    \label{fig:time_evolution}
\end{figure*}

\begin{figure*}
\includegraphics[width=2\columnwidth]{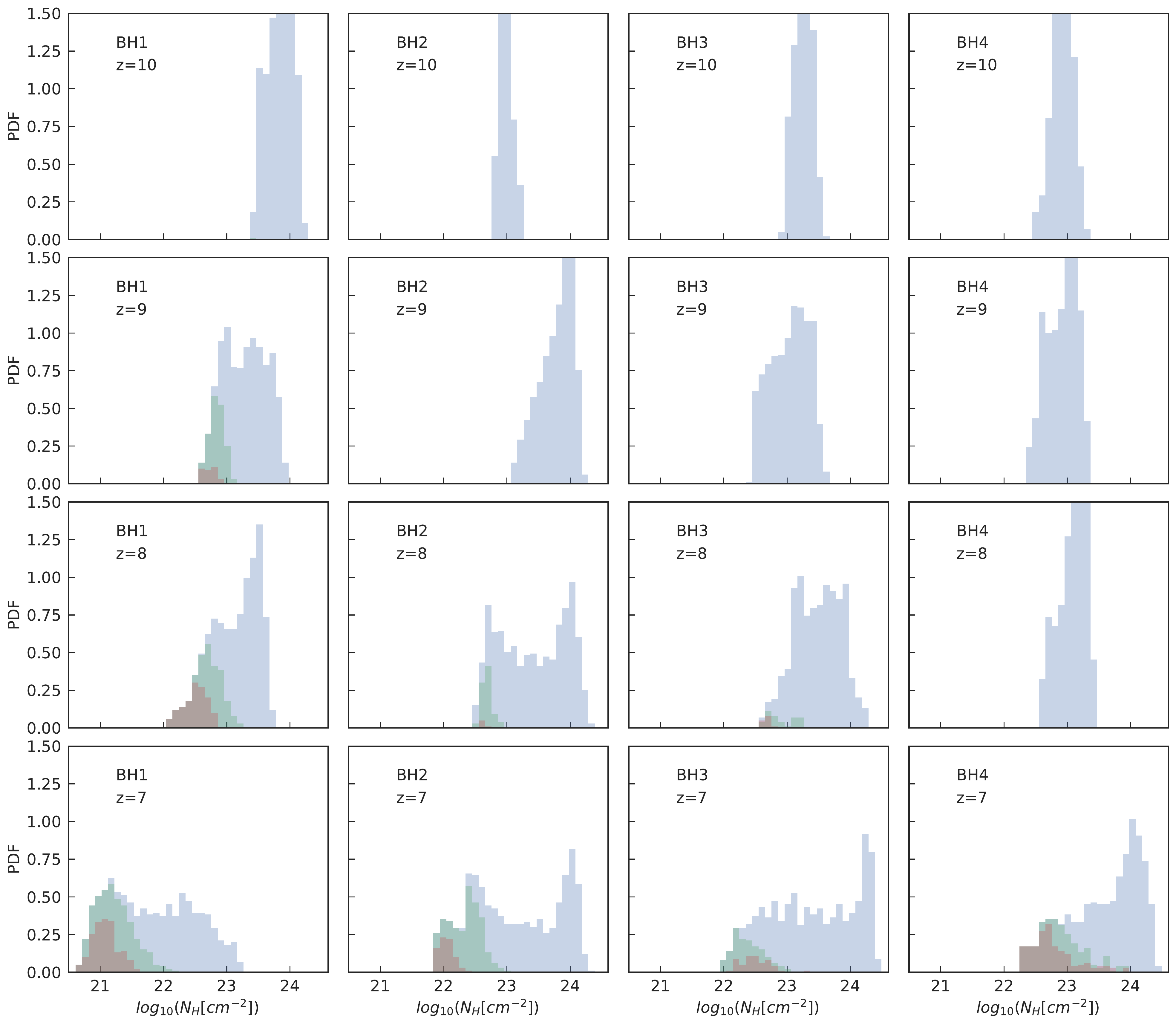}
    \caption{$N_{\rm H}$ histogram of the four sample QSOs from $z=10$ to $z=7$. The green shade marks out the contribution of lines of sight with $N_{\rm out}/N_{\rm H}>0.01$ while the red shade marks out the distribution of lines of sight with $v_r > 300$ km/s. All the panels share the same $x$ and $y$ axis.
    }
    \label{fig:sample4_histogram}
\end{figure*}

\begin{figure*}
\includegraphics[width=2.15\columnwidth]{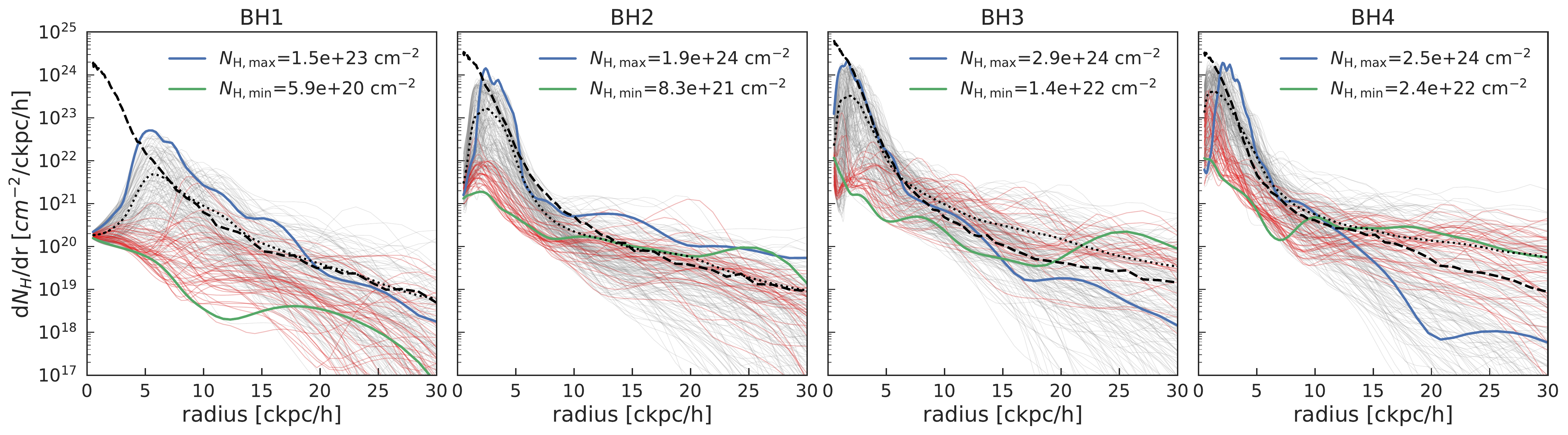}
    \caption{ Radial profiles of $N_{\rm H}$ around QSOs.
    The grey lines are the differential $N_{\rm H}$ profiles for 192 lines of sight around each sample QSO at $z=7.0$. 
    For each panel, the blue solid line gives the profile of the most obscured line of sight (i.e., that with the largest $N_{\rm H}$ value), and the green solid line corresponds to the least obscured line of sight. The black dotted lines mark out the mean value of all lines of sight at the corresponding radius.
    The radial profile for directions with strong outflows ($v_r>300$ km/s) are coloured pink. The dashed black lines in each panel trace  the corresponding stellar density profiles, converted into the appropriate units: $n_{\rm H} /[\rm cm^{2} \times \rm ckpc/h]$.
    }
    \label{fig:radial_profile}
\end{figure*}

% \begin{figure*}
% \includegraphics[width=2.1\columnwidth]{Ang_corr_4samples.pdf}
%     \caption{For each of the four samples, the profiles give the probability of a pair of line of sight both satisfied the given outflow criteria as a function of the separation angle.
%     }
%     \label{fig:Ang_corr_4samples}
% \end{figure*}

% Fig --------------------------------------------------------
% --------------------------------------

% 3.1 ------------------------------------------------------------------------------------------------
\subsection{Global BH properties of BlueTides}

In Figure~\ref{fig:Muv_collection} we show some basic properties of the bright AGN population in the \textsc{BlueTides} simulation from $z=10$ to $z=7$.
For each redshift, we select AGN with luminosity $L_X > 10^{43}$ ergs/s (corresponding to $L_{\rm bol} > 2 \times 10^{44}$ ergs/s). 
The top panels are the 2D histogram of the BH mass and luminosity. 
We compute the quasar luminosity in the X-ray band and show it on the upper $x$ axis and in the rest-frame UV-band along the  lower $x$ axis (see Section \ref{subsection:subsection 2.2}).
The brown dotted lines in the top panels of Figure~\ref{fig:Muv_collection} indicate twice the Eddington luminosity,
determined from the BH mass (see Eq.~\ref{equation:Meddington}). 
BHs lying on the brown dotted lines are in a state at the upper limit of allowed super-Eddington accretion.
The bottom panels of Figure~\ref{fig:Muv_collection} show the rest-frame UV band (intrinsic) luminosity of AGN compared to that of their host galaxies, with the black dashed lines indicating where galaxy and AGN are equally luminous in  the UV band. 
BHs below the black dashed lines are the quasars that have intrinsic UV luminosity that can outshine their host galaxy (without considering the dust attenuation).
When evolving to lower redshift, there are more QSOs that can outshine their host galaxy, most of them having high luminosity, $M_{\mathrm {UV}}<-22$ (or $L_X > 10^{43.5}$ ergs/s).

For illustrative purposes, we choose four sample QSOs to study their host environment, obscuration state, as well as their time evolution.
The objects are marked by colored stars in Figure~\ref{fig:Muv_collection}.
The basic properties of the 4 sample QSOs at $z=7$ are listed in Table $1$, with the host mass properties ($M_*$,$M_{\rm gas}$ and $M_{\rm H_2}$) all being calculated within the virial radius of the halo.
All  4 sample QSOs are massive SMBH with $M_{\rm BH} > 10^8 M_{\odot}$ and have UV band luminosities brighter than their host galaxies at $z=7$.
BH1 (in red) is the most massive SMBH in the simulation,  having grown to $M_{\rm {BH}} = 7.7 \times 10^8 M_{\odot}$ at $z=7$. We have studied its host galaxy properties and gas outflows in previous publications \citep{Tenneti2018,Ni2018}.
% BH2 (in orange) and BH3 (in green) are the second and the third most massive SMBH at $z=7$, BH3 is also the most luminous quasar at $z=7$. % BH4 (in purple) is also a typical bright quasar which is currently at Eddington accretion state at $z=7$.
 
To allow the reader to briefly examine the 4 QSO hosts, we plot the gas in their local environments as well as
images of their host galaxies in Figure~\ref{fig:images}.
For each  of the panels in Figure~\ref{fig:images}, the BH is shown in the center (position
marked by the red cross in top panels).
The top panels are the gas density field color coded by temperature (blue to red indicating cold to hot respectively), and bottom panels are stellar density field color coded by the age of stars (from blue to red indicating young to old populations respectively).
In each case, the BH is in the densest part of the gas distribution, surrounded by clumps and filaments of dense gas. 
The host galaxies vary in their morphology and in the spatial distribution of stellar ages.
We study the obscuration caused by their surrounding accreting gas and host galaxy in Section~\ref{subsection:QSOsamples}.
% 3.1  ------------------------------------------------------------------------------------------------

% 3.2 ------------------------------------------------------------------------------------------------
\subsection{Gas properties surrounding the QSOs}
\label{subsection:QSOsamples}

In this section, we take our four sample QSOs and explore the typical angular variation and time evolution of the column density $N_{\rm H}$, and study its relationship with dust extinction and gas outflow due to AGN feedback. 

In Figure~\ref{fig:BH0map}, we focus on BH1 and show the column density, dust optical depth and radial velocity of its surrounding gas.
The three gas properties are calculated along all lines of sight centered at BH1 and are plotted in Aitoff projection to illustrate the angular distribution.
Each row of Figure~\ref{fig:BH0map} represents the state at certain redshift so that we are tracing the time evolution from $z=10$ to $z=7$.
The leftmost column shows the hydrogen column density $N_{\rm H}$, from blue to red indicating low $N_{\rm H}$ to high $N_{\rm H}$.
The middle column gives the corresponding dust optical depth $\tau_{\rm UV}$ calculated based on Eq.~\ref{equation:tau_dust}, with blue to red representing low to high dust attenuation in this case.
In the third column we show the averaged radial velocity $v_r$ along the corresponding line of sight. Note that we define the positive values as being in the outward direction and negative values as the inward direction. This means that the yellow to red patches in the third column represent the regions where gas is flowing outward rapidly.

The middle column of dust optical depth $\tau_{\rm UV}$ exhibits a similar angular pattern to the $N_{\rm H}$ map, indicating that directions with larger $N_{\rm H}$ are also more likely to have higher dust extinction. In other words, dust attenuation of AGN mainly traces the regions with high gas density, and the gas metallicity $Z$ only modulates the variation at a sub-dominant level.

A comparison between the first and the third columns in Figure~\ref{fig:BH0map} indicates a clear correlation between the low column density regions and directions with high outward velocity.
This is consistent with our previous study of quasar-driven outflows \citep{Ni2018}, where we found that the outflowing gas tends to channel through low density regions. 
The underlying picture is that massive bright quasars were mostly enshrouded by high density gas, and as the quasars dumped feedback energy into the surrounding gas and drove outflows, this process opened up large regions of low $N_{\rm H}$ where outflows could pass through, opening a window for observations of the central object.

Maps of the other three sample QSOs show a similar pattern relating $N_{\rm H}$, $\tau_{\rm UV}$ and $v_r$, therefore we only plot the $N_{\rm H}$ field of the 3 samples in Figure~\ref{fig:3Nhmap}, and put the full 3-field maps of BH2, BH3 and BH4 in Appendix~\ref{sec:Appendix} as further illustration.
We study the relationship  of $N_{\rm H}$ with gas outflow and $\tau_{\rm UV}$ on a statistical basis in the next section. 
To further quantify the $N_{\rm H}$ distribution, we plot in Figure~\ref{fig:sample4_histogram} the $N_{\rm H}$ histograms of the four sample QSOs and mark the contributions from lines of sight with large outward radial velocity $v_r$ > 300 km/s in a red color. 
Another way to determine the outflow direction is to find in which lines of sight the outflowing gas particles reside. 
As established in \cite{Ni2018}, we find that outflow gas can be defined based on the criterion that a gas particle has a large enough velocity to escape from the potential well of the halo. i.e.,  peculiar velocity larger than the escape velocity $v_{\rm esc}$:
\begin{equation}
\label{equation:outflow}
    \frac{1}{2}m v_{\rm esc}(r)^2 \geqslant \int_{r}^{R_{200}} \frac{G M(<r') m}{r'^2} dr' +  \frac{G M(<R_{200}) m}{R_{200}}
\end{equation}
The green shaded regions in Figure~\ref{fig:sample4_histogram} mark the lines of sight where the outflow gas constitutes a nonegligible fraction of the overall column density,  $N_{\rm out}/N_{\rm H}>0.01$.
Here $N_{\rm out}$ is calculated based on the same formalism as Eq.~\ref{equation:NH}, but only considering the contribution from outflow gas when calculating Eq.~\ref{equation:rho_i}.
We see that gas under both criteria occupies the lower $N_{\rm H}$ region in the overall distribution, indicating a correlation between outflow and $N_{\rm H}$.

%%%%%%%%%%%%%%%%%%%%%%%%%%%%%%%%%% time evolution  %%%%%%%%%%%%%%%%%%%%%%%%%%%%%%%%%%
\subsubsection{Time evolution}

Both Figure~\ref{fig:Nhmaps} and Figure~\ref{fig:sample4_histogram} show a rapid evolution of $N_{\rm H}$ with redshift. 
BH1 is originally heavily obscured at $z=10$ with $N_\mathrm{H} \sim 10^{24} \mathrm{cm}^{-2}$, then with the launching of powerful feedback it clears out its high density gas environment and ends up with $N_\mathrm{H} \sim 10^{22} \mathrm{cm}^{-2}$ at $z=7$.
On the other hand, BH2, BH3 and BH4 all start with low column density ($N_{\rm H} < 10^{23} \rm{cm}^{-2}$ at $z=10$), but get progressively more obscured while also having increasing angular variations with time.

As illustrated by Figure~\ref{fig:sample4_histogram}, all the 4 samples had quite a uniform $N_{\rm H}$ distribution at $z=10$, with the difference between highest and lowest $N_{\rm H}$ smaller than 1 dex. 
The angular variation of $N_{\rm H}$ gets larger when going to lower redshift accompanied by the emergence of the outflow.
At $z=7$, the distribution of $N_{\rm H}$ with respect to different lines of sight can span over 2 dex.

More quantitatively, in Figure~\ref{fig:time_evolution} we show the time evolution of the four sample QSOs with regard to their BH mass, luminosity, Eddington accretion ratio and obscuration state.
We quantify the level of the obscuration around the QSO using $N_{\rm H,ave}$, which is the averaged value of $N_{\rm H}$ over all lines of sight (the fourth panel). 
To describe the angular variation in $N_{\rm H}$, we calculate the standard deviation of $N_{\rm H}$ based on all lines of sight $\sigma(\rm log_{10} N_{\rm H})$ (the fifth panel). 

Tracing the time evolution, we find that the four sample QSOs  have all reached a high accretion state with $L/L_{\mathrm{Edd}} \gtrsim 1$ during their evolutionary history (BH1 at $z=10$, BH2 at $z=9$, BH3 and BH4 at $z=7.5$). 
At that epoch, they were generally at a higher level of obscuration, with $N_{\rm H,ave} \sim 10^{24} \mathrm{cm}^{-2}$.
Figure~\ref{fig:time_evolution} also shows that massive BHs are unlikely to stay in a state of close to Eddington accretion, since the strong AGN feedback driven by the high luminosity will self modulate the surrounding gas density field, launching strong gas outflows that clear out part of the high density accreting gas (which has large obscuration), and increases the angular variations in the $N_{\rm H}$ field.

We study the relationship between angular variations of the obscuration and  QSO properties using larger,  statistical samples in Section~\ref{subsection:statistics}.

%%%%%%%%%%%%%%%%%%%%%%%%%%%%%%%%% radial contribution %%%%%%%%%%%%%%%%%%%%%%%%%%%%%
\subsubsection{Radial distribution}

Studying the length scales that make the highest contribution to galactic obscuration can give us more physical insight.
To do this we use the differential $N_{\rm H}$ profile $dN_{\rm H}/dr$, which shows 
the radial distribution of $N_{\rm H}$ along each line of sight.
In Figure~\ref{fig:radial_profile}, we select 192 evenly distributed lines of sight (using \textsc{Healpy}) for the 4 sample QSOs at $z=7$ and plot the corresponding differential $N_{\rm H}$ profiles as grey lines.
The $x$ axis indicates the distance (in co-moving coordinates) from the central QSO. 
Integration of the profile over $r$ gives the column density $N_{\rm H}$ of the corresponding line of sight.
For each panel, the blue/green line marks the most/least obscured line of sight, with  corresponding $N_{\rm H}$ given in the legends.
Comparing the blue and green lines we can see that the galactic obscuration surrounding a certain QSO can vary by over 2-3 orders of magnitude between different lines of sight, because of spatial variations in the density field.

The differential $N_{\rm H}$ radial profiles show clearly that for sightlines with large column density, the $N_{\rm H}$ is mostly contributed by high density gas clumps (e.g., those with $dN_{\rm H}/dr>10^{22}$ $[\rm cm^{-2}/\rm ckpc/h]$) which mostly reside within $r<10$ ckpc/$h$ (corresponding to $r < 1.8$ kpc in physical units) of the QSO.
More quantitatively, using the mean $N_{\rm H}$ values (black dotted lines), we find that $N_{\rm H} (<10 \rm{ckpc}/h)$ $\gtrsim 90\%$ $N_{\rm H} (<30 \rm{ckpc}/h)$.

The pink lines in each panel denote the lines of sight with strongly outflowing gas  ($v_r > 300$ km/s) for the sample QSOs. we can see that they mostly channel through the directions without high density clumps and indeed have low $N_{\rm H}$ values compared to the overall populations.

We would also like to compare the radial $N_{\rm H}$ distribution to the stellar density distribution of the host galaxy.
The black dashed lines in each panel of Figure~\ref{fig:radial_profile} show the averaged stellar density profile converted to 
an equivalent hydrogen number density in units of $n_{\rm H} /[\rm cm^{2} \times \rm ckpc/h]$.
(To convert to hydrogen number density, we divide the mean stellar density by the hydrogen mass $m_p$.)
% \rupert{how is this done? Dividing by the mean stellar density and  multiplying by the mean $N_{\rm H}$ ? Need to give details.}
Since the star forming gas in our model is the source of obscuration, it is straightforward to understand why the radial profile of the stellar density approximately traces the $N_{\rm H}$ profile, indicating that most of the obscuring gas resides in the host galaxy.

We emphasis that the $N_{\rm H}$ radial profile in Figure~\ref{fig:radial_profile} clearly shows that, for $z=7$ QSO, the host interstellar medium (ISM) is able to produce absorption up to Compton-thick level ($N_{\rm H} > 10^{24} \rm cm^{-2}$) on scale of a few ckpc. 
This is due to the ISM at these high redshift being much denser than that around the AGN at low redshift ($z \lsim 2$), where it is claimed that Compton-thick absorption can only be produced by the parsec scale gas/torus in the nuclear region \citep{Buchner2017-II}.

% host ISM is able to produce absorption up to Compton-thick (>~1e24 cm-2). This is indeed different 
% from low-z, where it has been claimed that Compton-thick absorption can only be produced by pc-scale gas, whereas the host ISM is always Compton-thin}

%% file: Sec3_Result_P2.tex
% 3.3 ------------------------------------------------------------------------------------------------

\begin{figure*}
\includegraphics[width=2\columnwidth]{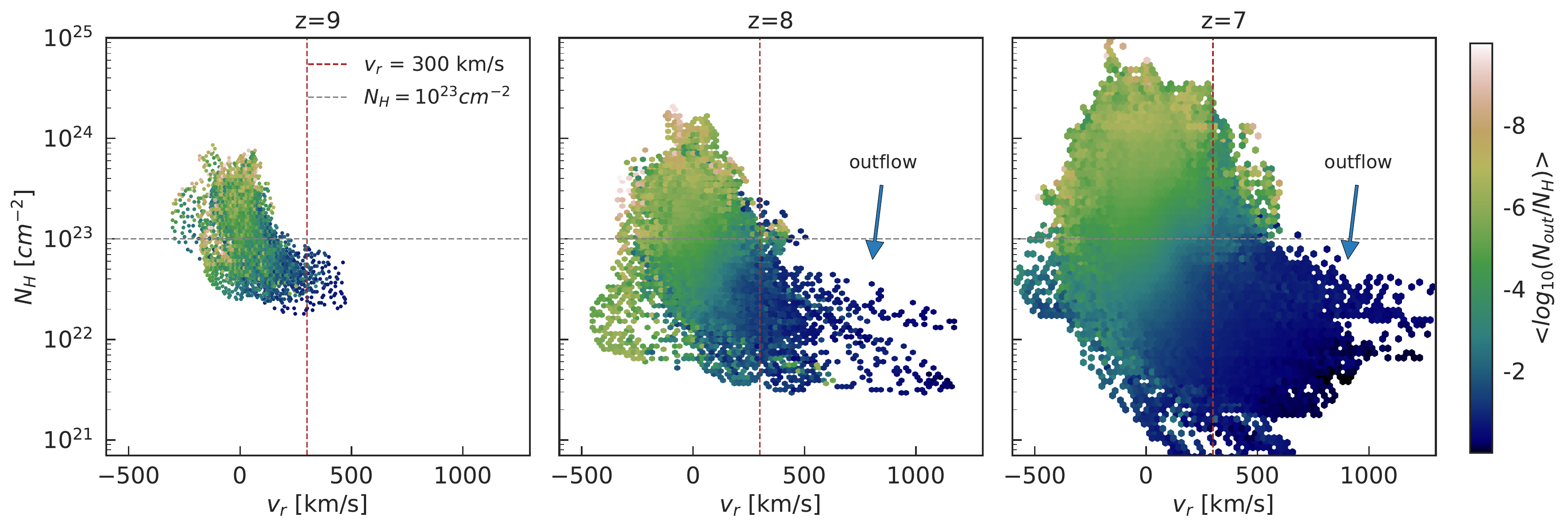}
    \caption{
  $N_{\rm H}$ versus radial velocity $v_r$ for all lines of sight around the QSO population with $L_X>10^{43}$ ergs/s. We show results at redshifts from $z=9$ to $z=7$. 
    The color coding indicates the averaged outflow fraction $N_{\rm out}/N_{\rm H}$ in each bin. The red vertical dashed line shows  where $v_r = 300$ km/s, and the grey horizontal dashed line is where $N_{\rm H} = 10^{23} \rm{cm}^{-2}$.
    }
    \label{fig:NH-vr-hist}
\end{figure*}

\begin{figure*}
\includegraphics[width=2\columnwidth]{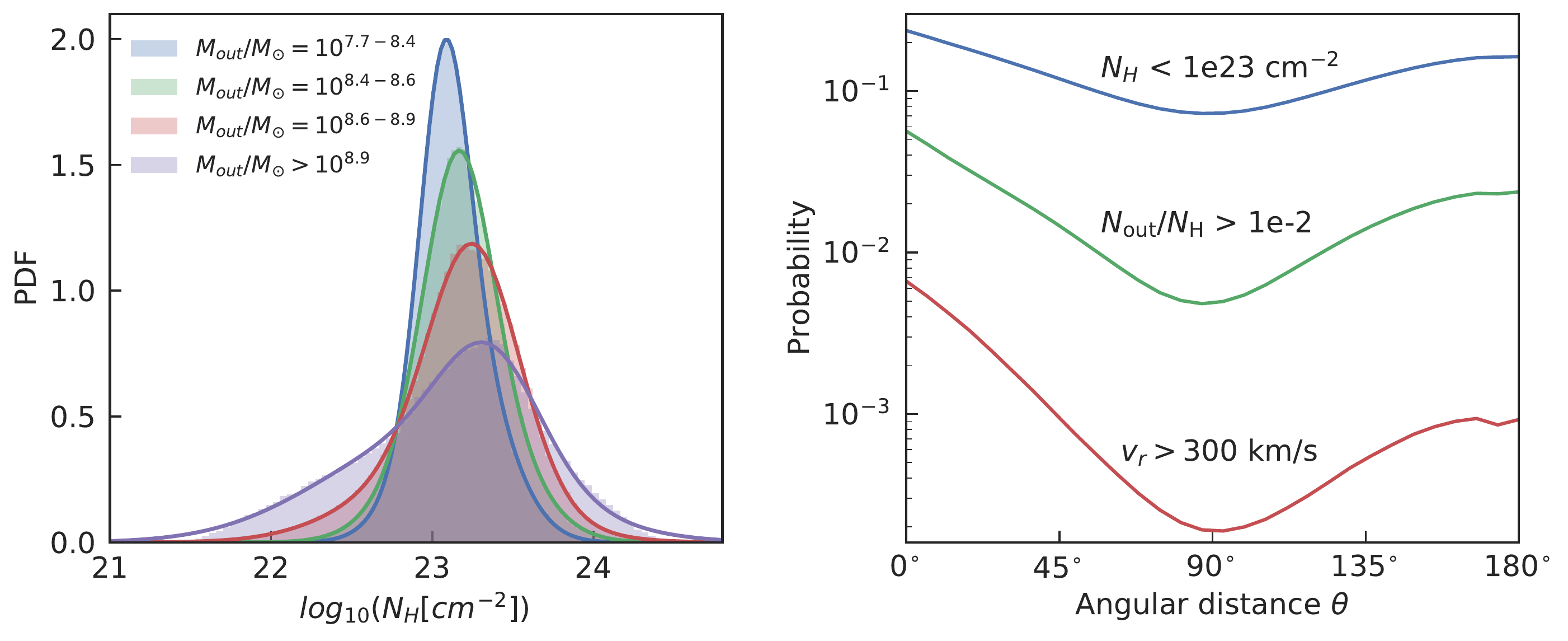}
    \caption{
    \textit{Left panel}: We split the $L_X > 10^{43}$ ergs/s QSO population at $z=7$ into different bins of mass of outflowing gas and plot the probability distribution function (PDF) of $N_{\rm H}$ calculated based on all lines of sight for QSOs in each bin. 
    \textit{Right panel:} The morphology of the gas outflow quantified using  angular correlations. 
    The profiles give the probability of both members of a pair of lines of sight satisfying the given criteria as a function of  separation angle. 
    The profiles are calculated by averaging over the QSO population with $L_X>10^{43}$ ergs/s at $z=7$ (see text for more detailed description).
    }
    \label{fig:ang_prob}
\end{figure*}

% Fig --------------------------------------------------------
% --------------------------------------

\subsection{Statistics of quasar obscuration}
\label{subsection:statistics}

In this section we study the obscuration state of the bright AGN population on a statistical basis.
We mainly focus on the $z=7$ bright AGN population with $L_X > 10^{43}$ ergs/s (corresponding to $L_{\rm bol}>10^{44.3}$ ergs/s).
In the simulation we have overall 3504 AGNs with $L_X > 10^{43}$ ergs/s at $z=7$, and these are widely spread with regard to BH mass and accretion state, residing in all different kinds of hosts. 
We explore the relation between the $N_{\rm H}$ distribution and outflows as well as the AGN and galaxy properties. 
%We will also trace to see whether there is a clear redshift evolution of $N_{\rm H}$.

\subsubsection{The N$_{\rm H}$ distribution around quasars and its relation to outflows}

% ------------------------------ NH-vr 3 redshift ----------------------------------------
First, we explore the relation between the N$_{\rm H}$ distribution and the outflow driven by AGN feedback.
In Section~\ref{subsection:QSOsamples} (Figure~\ref{fig:Nhmaps},\ref{fig:sample4_histogram},\ref{fig:radial_profile}), we have illustrated that a  positive correlation exists between gas outflow and low $N_{\rm H}$ regions in the hosts of our four sample QSOs.
We now study this relation statistically with larger samples.

In Figure~\ref{fig:NH-vr-hist}, we plot the $N_{\rm H}$ versus $v_r$ distribution for all lines of sight around the $L_X > 10^{43}$ ergs/s QSO population at $z=9$, $z=8$ and $z=7$.
In particular, we have overall 259, 637 and 3504 such bright AGNs at these three redshifts, and we plot the $N_{\rm H}$ distribution based on 972 lines of sight for each AGN.
The color coding indicates the averaged value of outflow fraction $N_{\rm out}/N_{\rm H}$ in each bin. 

We see that the lines of sight with large outward radial velocity indeed correspond to lower $N_{\rm H}$ regions and they generally have large outflow fractions with $N_{\rm out}/N_{\rm H} > 0.01$ (blue color). 
At $z=7$, for lines of sight with $v_r > 300$ km/s, 87\% of them have $N_{\rm H} < 10^{23} \rm cm^{-2}$.
The overall $N_{\rm H}$ distribution at high redshift $z=9$ is narrower because of the lack of powerful gas outflows.
When going to lower redshift, the $N_{\rm H}$ distribution is broadened, accompanied by more gas outflows launched by AGN feedback.

% ---------------------------------------- z = 7 ----------------------------------------
We  now focus on the latest redshift $z=7$ and study the relationship between $N_{\rm H}$ and outflow based on the QSO population with $L_X>10^{43}$ ergs/s.
In the left panel of Figure~\ref{fig:ang_prob} we plot the $N_{\rm H}$ probability distributions obtained by separating the QSO hosts into multiple $M_{\rm out}$ bins, where $M_{\rm out}$ is the mass of the outflow gas calculated based on outflow criteria Eq.~\ref{equation:outflow}.
Each bin here contains at least 300 AGNs (with 972 lines of sight used for analysis of each AGN).
The overall $N_{\rm H}$ distribution skews towards the low $N_{\rm H}$ end when more gas is outflowing from the  host.
For QSO hosts that have $M_{\rm out} > 10^{8.9} M_{\odot}$ (purple line), the probability of having $N_{\rm H} < 10^{23} \rm cm^{-2}$ can reach  40\%.
% Powerful outflow are typically launched by large massive system, this explains why the $N_{\rm H}$ distribution gets broadened when the $M_*$ or $M_{\rm H_2}$ become more massive.

The Aitoff projection maps in Figure~\ref{fig:Nhmaps} illustrate how the  morphology of the outflow can be bimodal for our 4 sample QSOs.
To study a larger QSO population, we apply a method inspired by angular correlation analysis to quantify the morphologies of outflow and angular $N_{\rm H}$ variations. 
The results are shown in the right panel of Figure~\ref{fig:ang_prob}, and are computed as follows.
For each AGN, we take pairs of sightlines from the 972 available, and calculate the probability that both members of a pair of sightlines 
with a specific angular separation $\theta$ satisfy particular physical criteria.
We calculate the probability by counting the number of sightline pairs in each angular separation bin that satisfy the criteria and divide by the total number of pairs in the bin. Our sample is the AGN population with $L_X>10^{43}$ ergs/s at $z=7$.
The blue line in Figure~\ref{fig:ang_prob} shows results for the first criterion, the probability that both sightlines 
have $N_{\rm H} < 10^{23} \rm cm^{-2}$.
We see that the probability becomes smaller when the separation angle $\theta$ increases from $0^{\circ}$ to $90^{\circ}$, and becomes larger again when $\theta$ keeps growing from $90^{\circ}$ to $180^{\circ}$. 
This convex shape can be produced if the averaged $N_{\rm H}$ distribution is bimodal in angle. This is in fact what we have seen in the example maps in Figure~\ref{fig:Nhmaps}. 
This morphology points to most of the QSOs having bimodal outflows, with their signature showing up as angular deficits in the obscuration. 
The red and green lines show the corresponding probability that a pair of lines of sight have $N_{\rm out}/N_{\rm H} > 0.01$ and $v_r > 300$ km/s. They have lower amplitude than for the $N_{\rm H}$ criteria because they are more stringent (as also illustrated in Figure~\ref{fig:sample4_histogram}). 
They however yield the same convex feature with the minimum probability at $\theta = 180^{\circ}$, indicating that on an averaged basis, the outflows in our simulation are indeed consistent with a bimodal structure.

% Fig --------------------------------------------------------
% --------------------------------------

\begin{figure*}
\includegraphics[width=2.1\columnwidth]{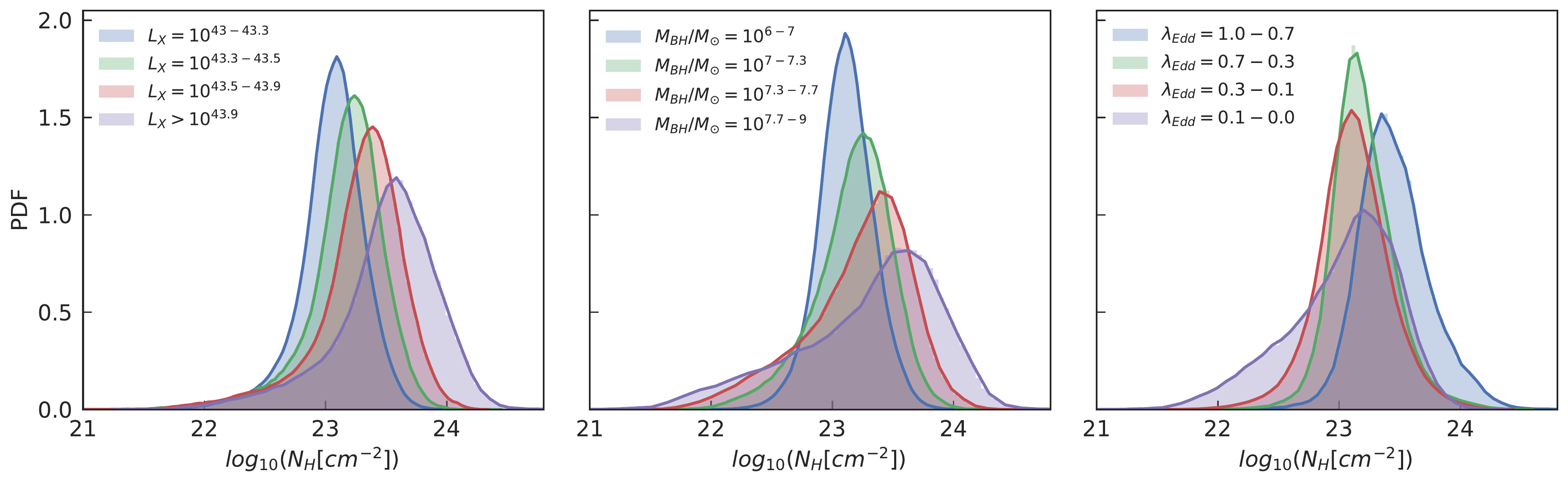}
    \caption{We split the $L_X > 10^{43}$ ergs/s QSO population at $z=7$ into different bins of luminosity (left), BH mass (middle) and Eddington accretion ratio (right) and study the corresponding probability distribution function (PDF) of $N_{\rm H}$ for all lines of sight in each bin. 
    The histograms are normalized so that the area under each curve is unity..
    }
    \label{fig:NH-hist-3bins}
\end{figure*}

\begin{figure}
\includegraphics[width=0.95\columnwidth]{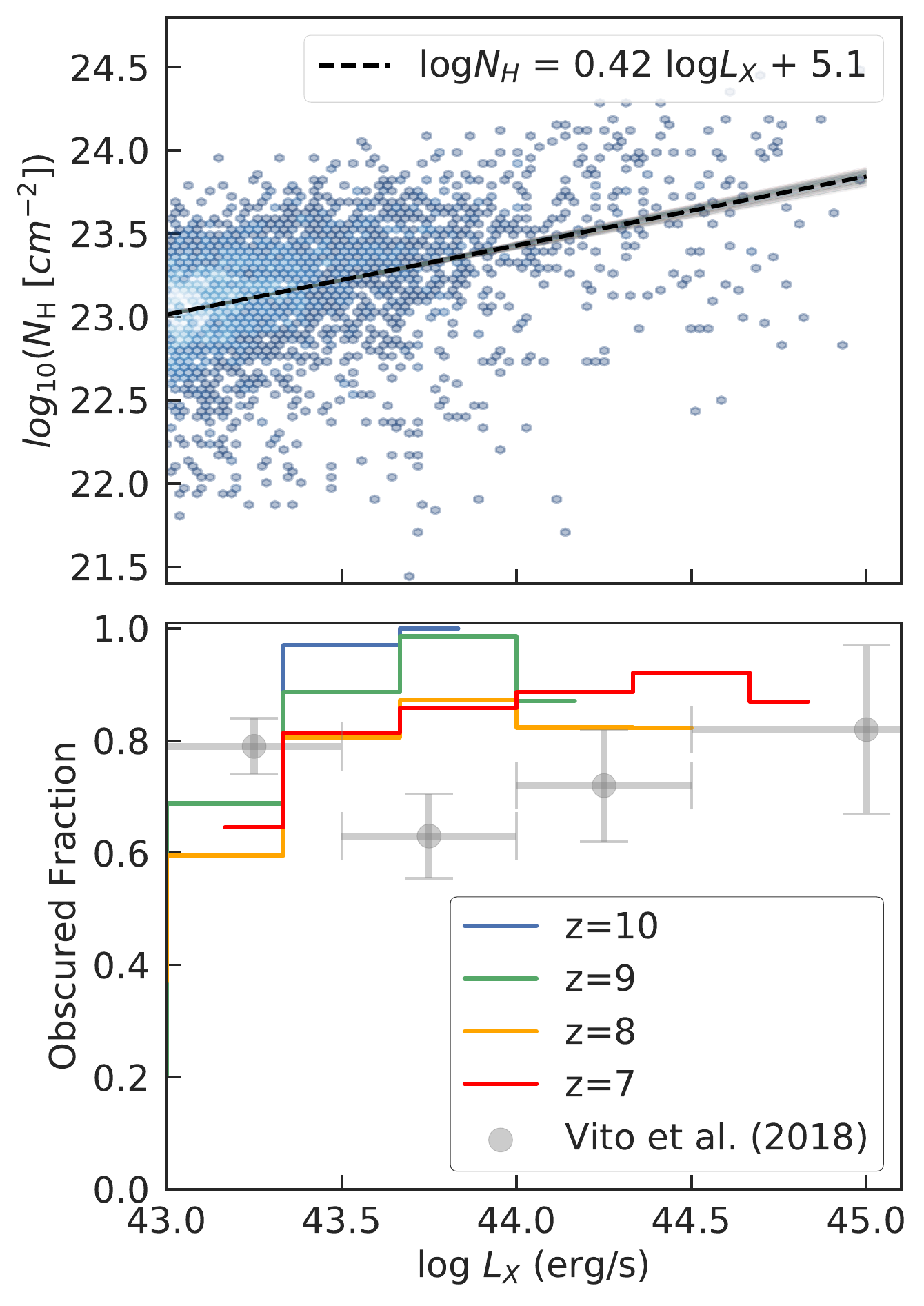}
    \caption{
    \textit{Upper panel}:  $N_{\rm H}$ vs. AGN luminosity for the  $L_X > 10^{43}$ ergs/s AGN population at z=7.
    The blue points show $N_{\rm H}$ along random lines of sight for AGN versus their intrinsic hard x-ray band luminosity $L_X$ 
    The black dashed line shows the average linear fit for all AGN, and the grey area overplots the linear fits made
     from all 972 realizations of $N_{\rm H}$ resulting from different lines of sight. 
    \textit{Bottom panel}:
    Binned estimates of the obscured AGN fraction (with $N_{\rm H}> 10^{23} \mathrm{cm}^{-2}$) as a function of $L_X$, shown from $z=10$ to $z=7$.
    The obscured fraction is calculated based on all lines of sight for the AGN populations in the corresponding luminosity bin. Grey data points with error bars are the observational results of \citet{Vito2018}, based on the AGN population at redshift $z$=3-6.
    }
    \label{fig:NH-Lx-fobsc}
\end{figure}

\begin{figure}
\includegraphics[width=0.95\columnwidth]{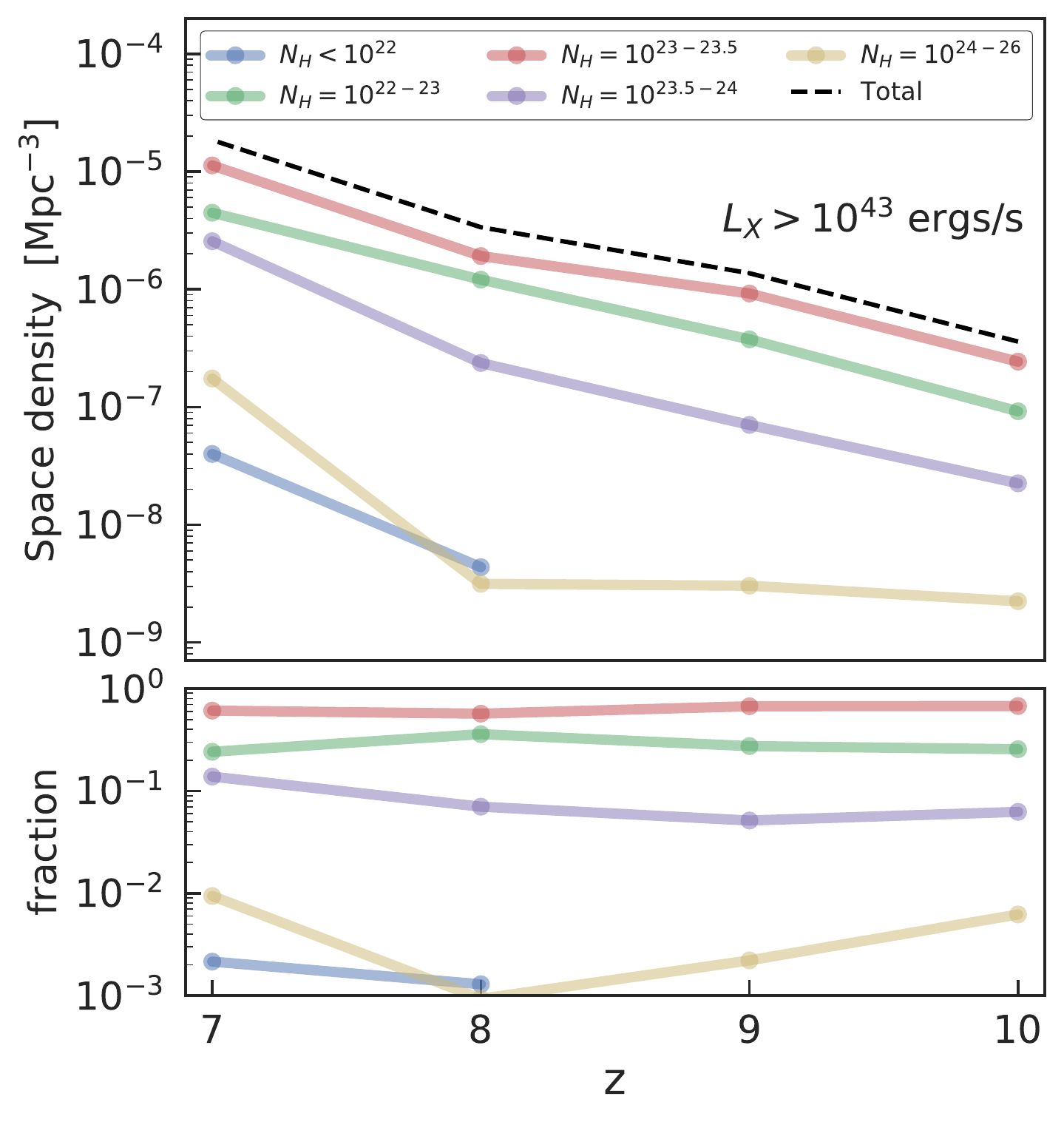}
    \caption{
    \textit{Upper panel}:
    The space density of the QSO population with $L_X > 10^{43}$ ergs/s as a function of the redshift. The black dashed line is the total space density. This is split into separate $N_{\rm H}$ bins (shown by coloured lines) by multiplying the fraction of the corresponding $N_{\rm H}$ component calculated based on all  QSO lines of sight.
    \textit{Lower panel}:
    The fraction of different $N_{\rm H}$ components, obtained by dividing the colored lines by the black dashed line in the upper panel.
    }
    \label{fig:Lx_space_density}
\end{figure}

% Fig --------------------------------------------------------
% --------------------------------------

%####################### NH-AGN relation ############################

\subsubsection{The relationship of $N_{\rm H}$ to AGN properties}

In this section, we explore the $N_{\rm H}$ distribution and its relationship to AGN properties, in particular, we are interested in how it varies with respect to AGN luminosity, BH mass and Eddington ratio.

In Figure~\ref{fig:NH-hist-3bins}, we split the QSO population with $L_X > 10^{43}$ ergs/s at $z=7$ into different bins of X-ray band luminosity $L_X$ (left panel), $M_{\rm BH}$ (middle panel) and the Eddington accretion ratio $\lambda_{\rm Edd}$ (right panel), and plot the overall probability distribution function (PDF) of $N_{\rm H}$ (including all lines of sight) for the corresponding bin.
We aim to show how the overall $N_{\rm H}$ distribution changes with respect to a few basic AGN properties. 
We note that each bin in Figure~\ref{fig:NH-hist-3bins} contains at least 300 AGNs (with 972 lines of sight used for each AGN). 

The left panel of Figure~\ref{fig:NH-hist-3bins} shows that $N_{\rm H}$ distribution becomes systematically broader for increasing AGN luminosity. 
The peak of the distribution changes from $N_{\rm H} = 10^{23.1} \rm cm^{-2}$ in the faintest bin (the blue line) to $N_{\rm H} = 10^{23.6} \rm cm^{-2}$ for the most luminous bin (purple). 
In addition, the $N_{\rm H}$ distribution develops a longer tail toward low $N_{\rm H}$ with increasing AGN luminosity. The lowest $N_{\rm H}$ lines-of-sight are created, as discussed in the previous section, as a result of the outflows driven by  AGN feedback. 
For the QSO population with $L_X > 10^{44}$ ergs/s (the purple line), the low $N_{\rm H}$ tail with $N_{\rm H} < 10^{23} \rm cm^{-2}$ constitutes 11\% of the overall distribution. 

The middle panel of Figure~\ref{fig:NH-hist-3bins} illustrates the relation between $N_{\rm H}$ and BH mass. It shows a similar trend with luminosity: the overall $N_{\rm H}$ distribution becomes broader and also skews toward lower $N_{\rm H}$ end with increasing BH mass. 
Quantitatively, the peaks of the $N_{\rm H}$ distribution as
a function of $M_{\rm BH}$ again moves from  $10^{23.1} \rm cm^{-2}$ to $10^{23.5} \rm cm^{-2}$ when going from the lowest BH mass bin (blue) to the highest mass bin (purple). 
The low $N_{\rm H}$ tail (with $N_{\rm H} < 10^{23} \rm cm^{-2}$) does however take up a larger fraction, 27\%, of the highest mass bin. 
This is again a consequence of the fact  that massive AGN are able to drive more powerful feedback which gives rise to the low $N_{\rm H}$ regions in their surroundings.
One fact to take into account is that the purple histograms in the left and middle panels both contain about 300 AGNs, however, the histogram in the middle panel is wider and more skewed, indicating that the $N_{\rm H}$ distribution for the most massive AGN population has larger variations compared to those for the most luminous population.

The right panel of Figure~\ref{fig:NH-hist-3bins} displays the $N_{\rm H}$ distribution as a function of Eddington ratio, 
the ratio of  bolometric luminosity to Eddington luminosity (which is only a function of the BH mass via Eq.~\ref{equation:Meddington}).
There is no clear trend in the $N_{\rm H}$ distribution in this case. However, comparing the purple with the blue histogram shows that the QSO population with lower Eddington ratio is more likely to be less obscured. 
For QSOs with $\lambda_{\rm Edd} < 0.1$ (purple line), the probability of finding $N_{\rm H} < 10^{23} \rm cm^{-2}$ is 40\%, while for $\lambda_{\rm Edd} > 0.7$ the probability is only 4\%.
We note, however that the low $N_{\rm H}$ values for AGNs with low Eddington accretion are not necessarily driven by  outflows since low $\lambda_{\rm Edd}$ could also be caused by low luminosity, where the AGN is not surrounded by high density gas environment.

To further quantify the correlation between $N_{\rm H}$ and AGN luminosity, we fit a linear relation between log$N_{H}$ and log$L_{X}$. 
As illustrated in Figure~\ref{fig:NH-hist-3bins}, the scatter in $N_{\rm H}$ is larger for increasing  luminosity. 
Therefore we use an up-sampling method to exploit the underlying angular distribution of $N_{\rm H}$ for each AGN, and compute uncertainties on the regression parameters.
We take the $N_{\rm H}$ value observed from 1 line of sight (out of 972 lines of sight for each AGN) as one realization and carry out a linear regression of $N_{\rm H}$ versus AGN luminosity $L_X$ for all realizations. 
In the upper panel of Figure~\ref{fig:NH-Lx-fobsc} we plot the results from these fits with the grey lines for all the 972 realizations, and a black dashed line showing the averaged relation.
The linear regression gives:
\begin{equation}
    \log N_{\rm H} = (0.42 \pm 0.02) \log L_{X} + (5.1 \pm 0.7)
\end{equation}
The blue points in the upper panel show  $N_{\rm H}$ versus $L_X$, for a single realization. 
The intrinsic scatter in the $N_{\rm H}$ versus $L_X$ relation for a single realization is approximately 0.5 dex.

Similar analysis can be used to fit a linear relation to  $\log N_{\rm H}$ and $\log M_{\rm BH}$. 
For the QSO population with $L_X > 10^{43}$ ergs/s at $z=7$, we get 
$\log N_{\rm H} = (0.13 \pm 0.02) \log M_{\rm BH} + (22.3 \pm 0.2)$.
However, as discussed for Figure~\ref{fig:NH-hist-3bins}, the variance of  the $N_{H}$ distribution for the most massive AGN bin is much larger than that for the most luminous AGN population.
 The linear fit of log$N_{H}$ to log$M_{\rm BH}$ therefore has a larger scatter at the massive $M_{\rm BH}$ end than the low end in each of the realizations.

We now compare our results to current observational constraints on the obscured fraction of high-redshift QSOs.
In the lower panel of Figure~\ref{fig:NH-Lx-fobsc}, we calculate the obscured QSO fraction in \textsc{BlueTides}, defined as the probability of a line of sight having $N_{\rm H} > 10^{23} {\rm cm}^{-2}$. We do that as a function of the  AGN luminosity and plot using colored solid lines the 
results obtained from $z=10$ to $z=7$.  
% and make comparison with the current AGN observation results at lower redshift.
% The colored solid lines show the results
Grey data points with error bars are the results from \cite{Vito2018}, based on observations of AGN from $z=3$ to $z=6$.
Note that we apply the same criteria for "obscured fraction" as that used by \cite{Vito2018} for high-$z$ QSOs.
Those authors use $N_{\rm H} = 10^{23} \rm cm^{-2}$ as the minimum column density for (heavily) obscured AGN at high redshift. 
This choice is different from the typical definition of the unobscured column density threshold $N_{\rm H} = 10^{22} \rm cm^{-2}$ because of the limited spectral quality of high-redshift X-ray observations \citep[see discussions in][for more details]{Vito2018}.

As can be expected from our previous consideration of the left panel of Figure~\ref{fig:NH-hist-3bins}, there is a weak trend whereby the obscured fraction increases when luminosity increases from $L_X \sim 10^{43}$ to $L_X \sim 10^{44}$ ergs/s. 
When going to even higher luminosity ($L_X > 10^{44}$ ergs/s), the obscured fraction does not keep growing because of the strong feedback brought by the bright AGNs.
We note however that our simulation cannot resolve the parsec scale torus-like structure around the AGN. 
We could therefore be underestimating the number of sightlines for which an AGN would be obscured by the torus.
This could explain the lower obscured fraction at $L_X \sim 10^{43}$ ergs/s compared to the observational results, because the overall galactic $N_{\rm H}$ distribution of AGNs in this luminosity bin peaks very close to the threshold value $N_{\rm H} = 10^{23} \rm cm^{-2}$, as shown in the left panel of Figure~\ref{fig:NH-hist-3bins}.

Overall, we predict that the obscured fraction for the $L_X > 10^{43}$ ergs/s QSO population at $z \geq 7$ in our simulation ranges from 0.7 to 1, which is comparable with (or slightly higher than) the current observational results based on $z=3$ to $z=6$ QSOs.

% ####################### redshift evolution ############################

\subsubsection{Redshift evolution}

From the lower panel of Figure~\ref{fig:NH-Lx-fobsc}, we can see that there is no clear trend of evolving obscured fraction with  redshift. 
More explicitly, in Figure~\ref{fig:Lx_space_density} we give the redshift evolution of the space density of AGN with $L_X > 10^{43}$ ergs/s and split the total space density at certain redshifts into different $N_{\rm H}$ bins calculated based on all lines of sight for the corresponding AGN population.
The blue line represent the $N_{\rm H} < 10^{22} \rm cm^{-2}$ component which corresponds to the typical definition of unobscured QSOs. 
The green line represents $N_{\rm H}$ values from [$10^{22} -  10^{23} \rm cm^{-2}$], the 
red and purple lines $N_{\rm H}$ between [$10^{23} -  10^{24} \rm cm^{-2}$], and the yellow line shows the Compton-thick fraction with $N_{\rm H} > 10^{24} \rm cm^{-2}$.
We see that the $N_{\rm H}$ distribution is mainly dominated by column densities $N_{\rm H} \sim 10^{23-23.5} {\rm cm}^{-2}$, and the fraction in each $N_{\rm H}$ bin does not change much with redshift.
The same analysis based on the QSO population above a higher luminosity cut (e.g. $L_X > 10^{43.5}$ ergs/s, not shown in a Figure) shows the same evolution of $N_{\rm H}$ fraction.
We therefore conclude that the overall $N_{\rm H}$ distribution around QSOs above a certain luminosity threshold ($L_X > 10^{43}$ ergs/s) does not show any strong evolution with redshift from $z=10$ to $z=7$.

%% file: Sec3_Result_P3.tex
\begin{figure*}
\includegraphics[width=2.1\columnwidth]{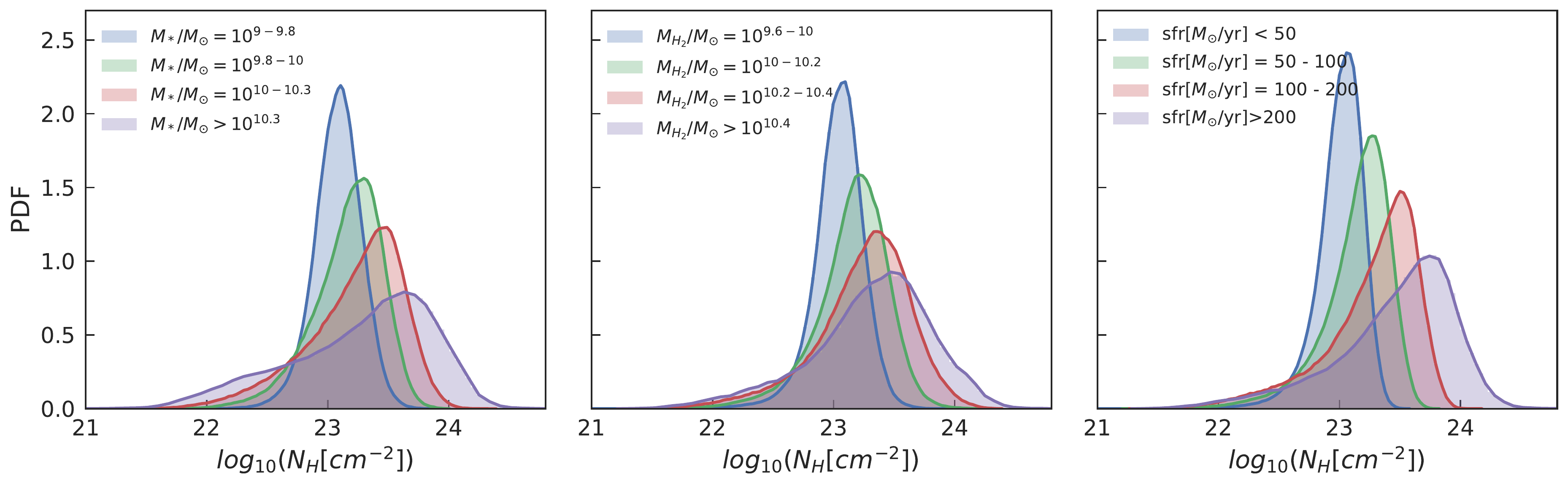}
    \caption{
    As in Figure~\ref{fig:NH-hist-3bins}, here we plot the overall $N_{\rm H}$ histogram by splitting the QSO population with respect to their host properties:  stellar mass (left), molecular mass (middle) and star formation rate (right).
    All the quantities are calculated within the virial radius $R_{200}$ of the QSO host.
    We include all QSOs with $L_X>10^{43}$ ergs/s at $z=7.0$.
    }
    \label{fig:NH_hist_hostbins}
\end{figure*}

\begin{figure}
\includegraphics[width=1\columnwidth]{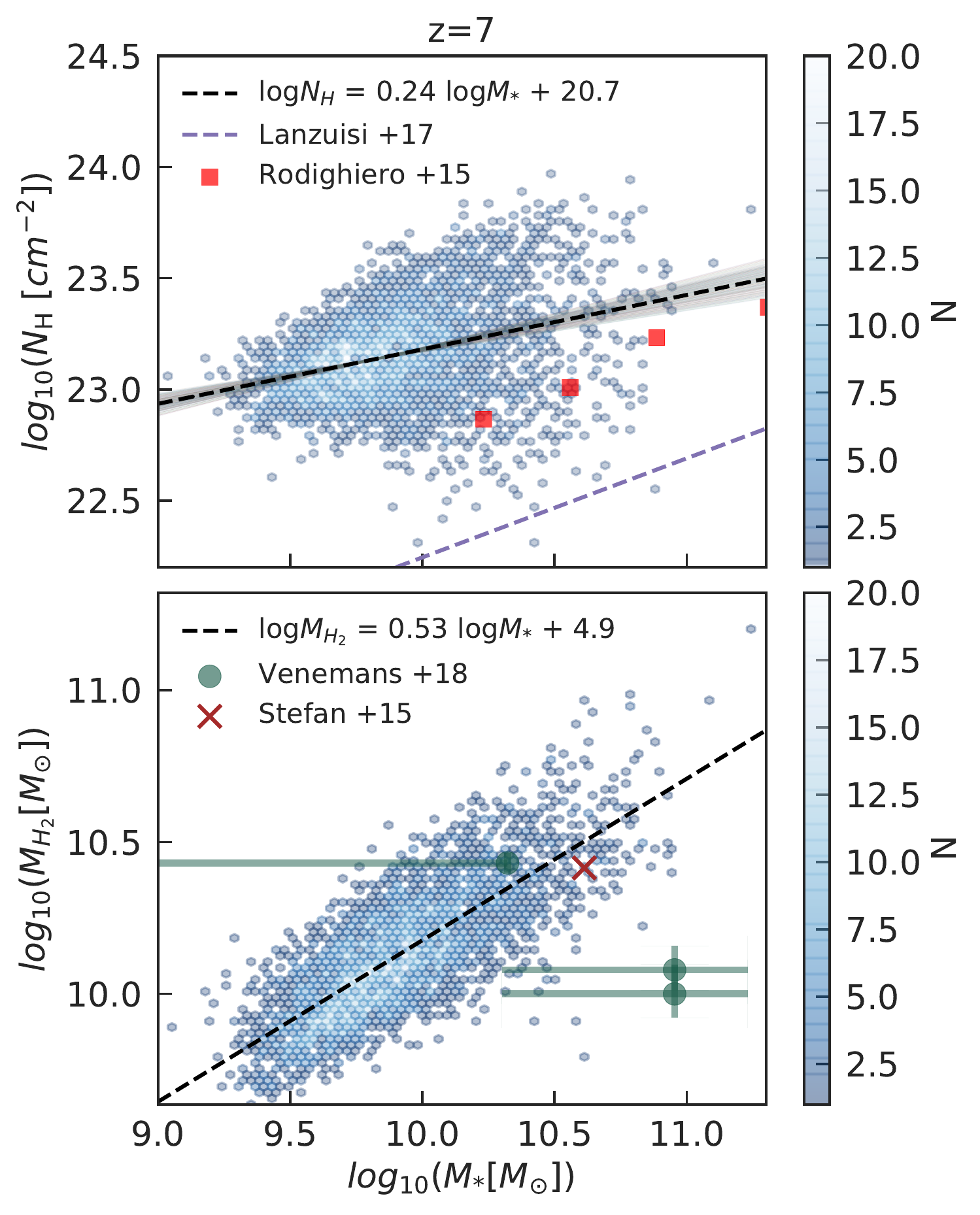}
    \caption{The gas properties of AGN versus the stellar mass of the host galaxy, for the QSO population with $L_X>10^{43}$ ergs/s at $z=7.0$.
    \textit{Upper panel}: The blue points give $N_{\rm H}$ for AGN versus the stellar mass of the host galaxy along a random line of sight. The grey areas show the linear fits for 972 realizations of $N_{\rm H}$ versus $M_*$ along different line of sight, and the black dashed line gives the averaged fitting value.
    The red squares show results from the observations of \citet{Rodighiero2015} at $z \sim 2$, and the purple dashed line shows the relation found by \citet{Lanzuisi2017} at $2<z<4$. 
    \textit{Lower panel}: The molecular mass in the host halo of AGN versus the stellar mass of the host galaxy. 
    The green points with error bars denote the observations of three quasar host galaxies at $6.6 < z < 6.9$ from \citet{Venemans2017}. 
    The brown cross represents the measurement from \citet{Stefan2015} for the $z=6.42$ host of QSO J1148+5251.
    }
    \label{fig:NH_MH2_Mstellar}
\end{figure}

% \begin{figure}
% \includegraphics[width=\columnwidth]{zbar_los.pdf}
%     \caption{
%     Purple shade give the 2D histogram of averaged metallicity along line of sight versus $N_{\mathrm H}$ for all AGN population with $L_X > 10^{43}$ ergs/s at $z=7.3$. Blue shade marks out the region covered by the most luminous population (19 AGNs with $L_{\rm bol} > 10^{46}$ ergs/s.)
%     Two horizontal lines mark out the level of solar metallicity ($Z_{\odot}=0.02$) and the SMC metallicity ($Z_{\rm SMC}=0.005$). 
%     The histogram is calculated based on all lines of sight, (i.e., 192 lines of sights for all 5077 AGNs with  $L_X > 10^{43}$ ergs/s at $z=7.3$.)
%     Here the averaged metallicity along the corresponding line of sight is calculated as $Z_{\mathrm{ave}=} = \int n_{\rm H} Z dl / \int n_{\rm H} dl$.
%     }
%     \label{fig:Z_NH}
% \end{figure}
% Fig ----------------------------------------------------------------------------------------------

% ----------------------------------------------------------------------------------------------

\subsubsection{Relationship of $N_{\rm H}$ to host galaxy properties}

% -------------------------------------------- M_* --------------------------------------------------

In this section we explore the relationship between the $N_{\rm H}$ distribution and QSO host properties.
In particular, we are interested in the variation of obscuration with stellar mass, molecular gas mass and the star formation rate in the host galaxy.

Similar to the style of Figure~\ref{fig:NH-hist-3bins}, in Figure~\ref{fig:NH_hist_hostbins} we split the QSO population with $L_{X} > 10^{43}$ ergs/s at $z=7$ into different bins of galaxy stellar mass $M_*$ (left panel), $M_{\rm {H_2}}$ (middle panel) and  star formation rate (right panel). We give the overall probability density distribution of $N_{\rm H}$ including all lines of sight for the AGNs in each bin.
Here all properties are calculated within the virial radius $R_{200}$ of the QSO host, and each bin contains at least 300 objects (with 972 lines of sight for each AGN).

If we consider the left panel of Figure~\ref{fig:NH_hist_hostbins} we can see that the overall $N_{\rm H}$ distribution peaks at higher $N_{\rm H}$ and becomes more broadened as we move to bins representing more massive $M_*$ values. 
For QSOs with $M_* > 10^{10.3} M_{\odot}$ (purple contour), the probability of finding $N_{\rm H} < 10^{23} \rm cm^{-2}$ is 29\%.
To quantify the relationship between $N_{\rm H}$ and $M_*$, we again use the up-sampling method described in the previous section and apply a linear regression analysis to all the realizations. 
The fit results are given in the upper panel of Figure~\ref{fig:NH_MH2_Mstellar}, with the grey lines being the fits for all  972 realizations, with the black dashed line the averaged value.
The linear fitting formula gives:
\begin{equation}
    \log N_{\rm H} = (0.24 \pm 0.03) \log M_{*} + (20.7 \pm 0.3)
\end{equation}
The blue points show the $N_{\rm H}$ from one random realization versus  $M_*$. From this we note that the intrinsic scatter is large ($\sim 0.7$ dex) at the massive end of $M_*$. 

Observations at lower redshifts also find a positive correlation between $N_{\rm H}$ and $M_*$.
For example, \cite{Lanzuisi2017} fit the linear relation between $N_{\rm H}$ and $M_*$ based on a sample of X-ray detected AGN and their far-UV detected host galaxies in the redshift range $0.1<z<4$, giving a slope in the range $\alpha = 0.42-0.88$ for different redshift bins.
The purple dashed line in Figure~\ref{fig:NH_MH2_Mstellar} show their linear regression result for the $2<z<4$ QSO population.
The red squares in Figure~\ref{fig:NH_MH2_Mstellar} show the result found by \cite{Rodighiero2015} for a sample of $z \sim 2$ AGN hosts in the COSMOS field. 

The above two observations probe $N_{\rm H}$ through AGN lines of sight. 
Additionally, \cite{Buchner2017} probe galactic obscuration through $N_{\rm H}$ values inferred from the X-ray spectra of GRBs, and study how $N_{\rm H}$ varies with respect to GRB host galaxy mass. 
These authors find that $N_{\rm H} \propto M_*^{0.38}$ in the redshift range $1<z<5$, implying that galactic obscuration scales with the galaxy size.
We note here, however that $N_{\rm H}$ see along AGN lines of sight is mainly contributed by the innermost regions of galaxies (< 10 ckpc/h) and is subject to large variation due to  AGN feedback. $N_{\rm H}$ therefore does not exhibit a tight power law relation with the $M_*$, because of the large angular variations present in the massive systems.

% -------------------------------------------- MH2 --------------------------------------------------

We would also like to quantify how $N_{\rm H}$ is related to  the overall gas mass in a host. 
We choose the mass of molecular gas mass as a proxy for gas mass because it is a direct observable, and also is the dominant contributor to the obscuration, since in high density regions the molecular fraction is close to 1.
In the middle panel of Figure~\ref{fig:NH_hist_hostbins} we split the overall $N_{\rm H}$ distribution in multiple $M_{\rm H_2}$ bins of QSO hosts.
Similar to the trend for $M_*$, we find a higher probability peak and larger $N_{\rm H}$ 
dispersion for QSOs with larger molecular gas mass. 
For QSOs with $M_{\rm H_2} > 10^{10.4} M_{\odot}$ (purple line), the probability of getting $N_{\rm H} < 10^{23} \rm cm^{-2}$ is 22\%.
Linear regression of $\log N_{\rm H}$ and $\log M_{\rm H_2}$ gives:
\begin{equation}
\log N_{\rm H} = (0.47 \pm 0.03) \log M_{\rm H_2} + (18.4 \pm 0.3)
\end{equation}

The lower panel of Figure~\ref{fig:NH_MH2_Mstellar} shows the mass of molecular gas $M_{\rm H_2}$ in the host versus $M_*$ for the same QSO population at z=7. 
There is a tight positive correlation between $M_*$ and $M_{\rm H_2}$ which can be fit by $\log M_{\rm H_2} = 0.53 \log M_{*} + 4.9$.
The coefficient 0.53, the slope of the mean relation between $M_*$ and $M_{\rm H_2}$, explains why the $\log N_{\rm H}$ slope with respect to $M_{\rm H_2}$ is steeper than that with $M_*$ ($0.53 \times 0.47 \sim 0.24$).
This indicates that the positive correlation between $N_{\rm H}$ and $M_*$ is essentially driven by the fact that larger galaxies are more gas enriched and therefore have more obscuring gas and dust along the lines of sight to their AGN. 
We now compare our molecular gas data with observations of high redshift QSO hosts.
The green data points with error bars in  Figure~\ref{fig:NH_MH2_Mstellar} are the observations of three QSO host galaxies residing at redshift $6.6 < z < 6.9$ from ALMA observations \citep{Venemans2017}.
The brown data point is the observation of a $z=6.42$ QSO host by \cite{Stefan2015}. To arrive at these points,
the galaxy (stellar) mass was inferred by subtracting the mass of molecular gas from the dynamical mass. 
We see that the molecular masses predicted by our simulation are broadly at the level seen in current observations of $z \sim 7$ AGN hosts, this provides a zeroth order sanity check on our estimate of galactic QSO obscuration.

% \yueying{0.46, 0.24, 0.53, These coefficients are very interesting!}

% -------------------------------------------- sfr --------------------------------------------------

Finally, we investigate the relation between $N_{\rm H}$ and the total star formation rate of the host galaxy. Here the star formation rate is calculated within the virial radius, $R_{200}$ of the halo.
Since the star formation rate correlates strongly with the local density field and the amount of molecular gas, it is not surprising to find a similar trend as that for $N_{\rm H}$ with $M_{\rm H_2}$.
As shown in the third panel of Figure~\ref{fig:NH_hist_hostbins}, we split the overall $N_{\rm H}$ distribution with respect to host star formation rate. 
It turns out that the $N_{\rm H}$ distribution systematically moves to higher values when we look at hosts with a larger star formation rate. 
At the same time, low $N_{\rm H}$ tail remains, due to the outflows driven by the massive systems. 
Quantitatively, for QSOs with sfr $ > 200 M_{\odot} / \rm yr$ (purple line), the probability of finding $N_{\rm H} < 10^{23} \rm cm^{-2}$ is 16\%.
Linear regression between $\log N_{\rm H}$ and the star formation rate gives:
\begin{equation}
\log N_{\rm H} = (0.48 \pm 0.02) \log (\rm sfr) + (22.31 \pm 0.04),
\end{equation}
where the star formation rate (sfr) is given in unit of $M_{\odot}/\rm yr$.

% ########################################## UVLF ############################################
% fig ----------------------------------------------------------------------------------------------
\begin{figure}
\includegraphics[width=1.05\columnwidth]{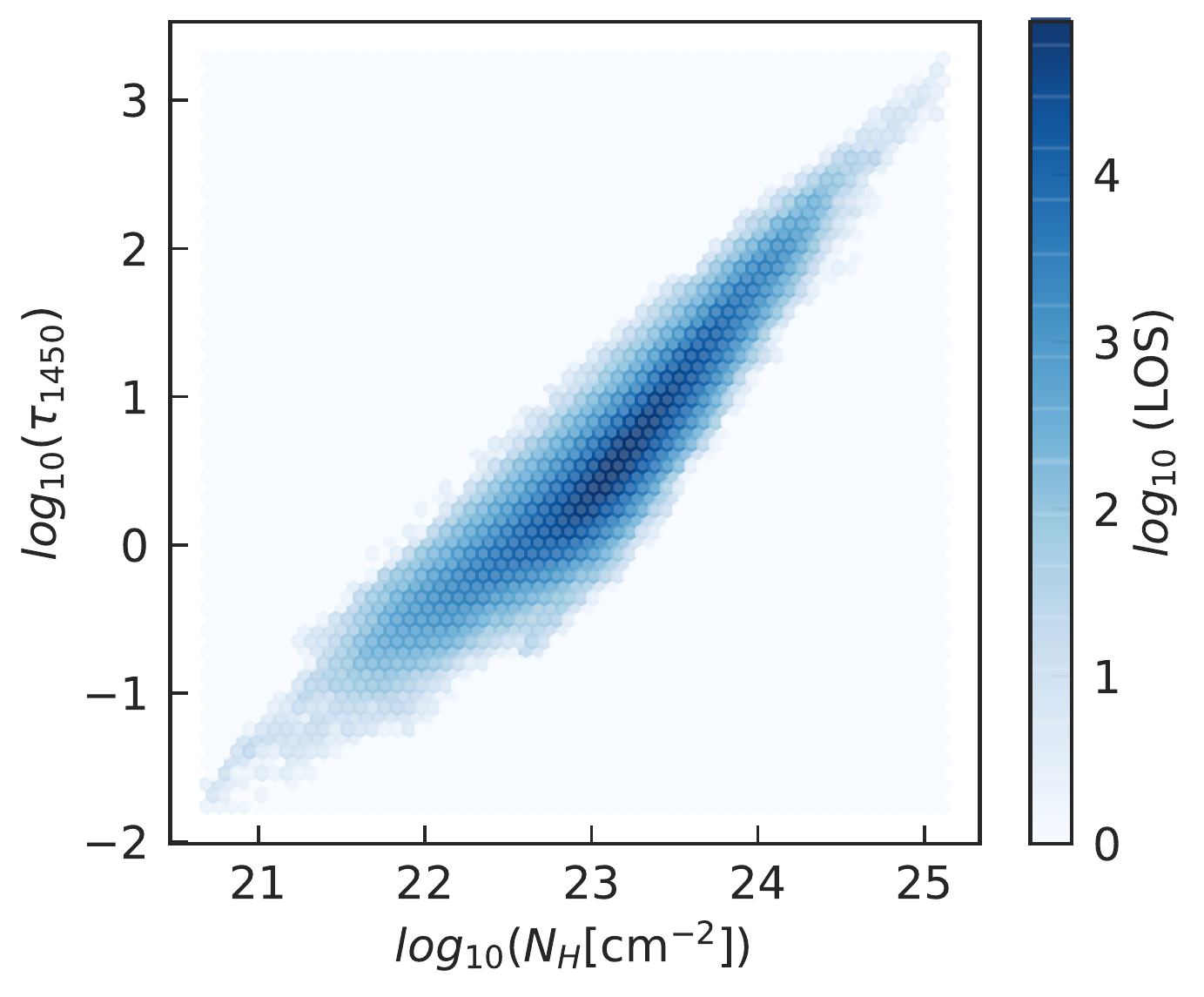}
    \caption{
    2D histogram of optical depth in the UV band ($\tau_{1450}$) due to dust grains versus the corresponding $N_{\rm H}$ for the entire AGN population with $L_X > 10^{43}$ ergs/s at $z=7$. 
    The histogram is based on all lines of sight, (i.e., 972 lines of sight for every AGN with  $L_X > 10^{43}$ ergs/s at $z=7.0$.)
    }
    \label{fig:tau_NH}
\end{figure}

\begin{figure}
\includegraphics[width=0.98\columnwidth]{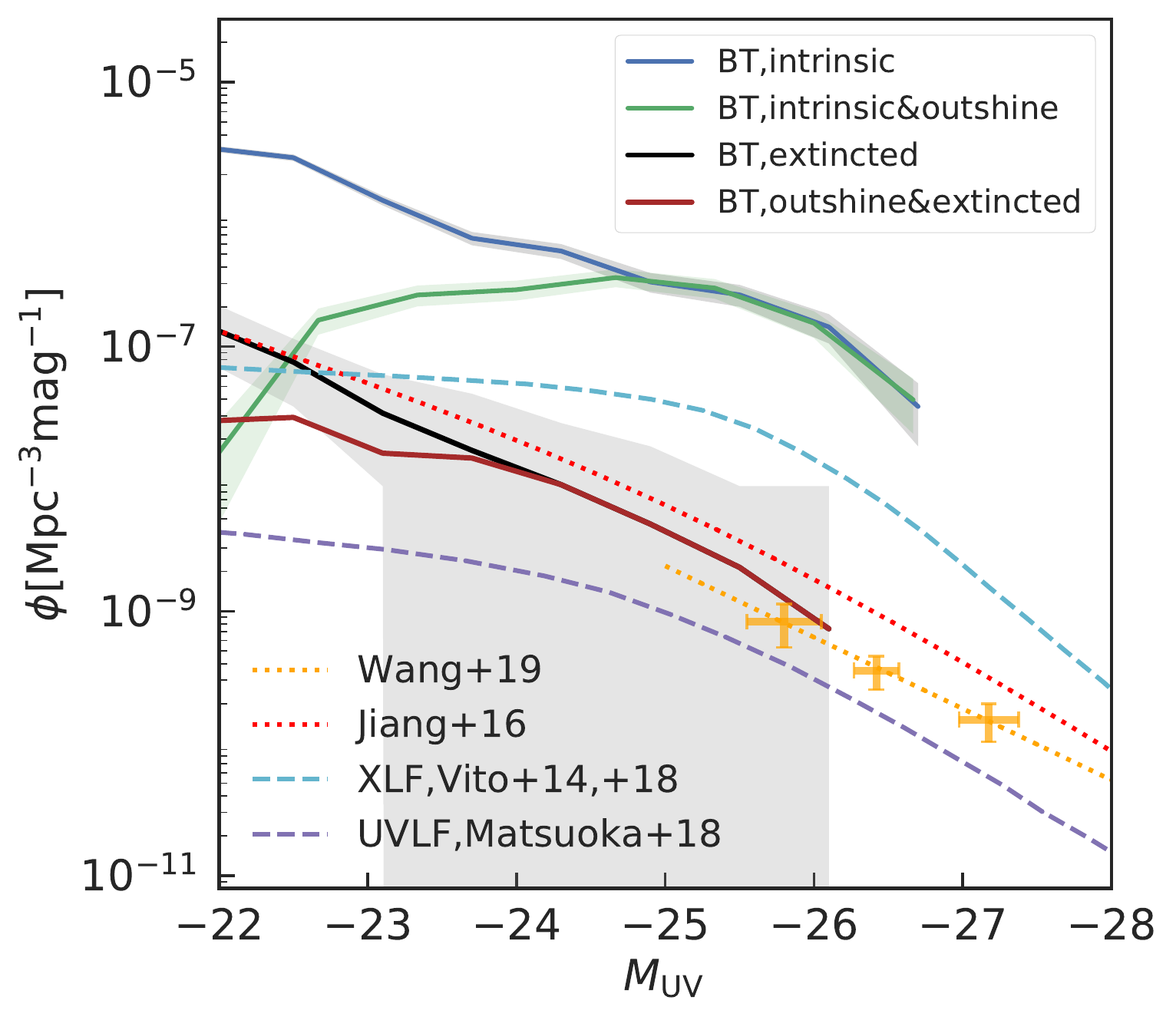}
    \caption{
    UVLF of QSOs. The blue solid line is the intrinsic UV band luminosity of QSOs predicted by \textsc{BlueTides} at $z=7.0$.
    The black solid line is the result taking dust obscuration into consideration.
    The grey shaded area gives the range of realizations with respect to different lines of sight (i.e., the result for the UVLF observed along 972 different directions).
    The orange solid symbols with error bars are the measured $z \sim 6.7$ UVLF from \citet{Wang2018}. The red dotted line is the $z \sim 6$ fitted UVLF measured by \citet{Jiang2016}. 
    The purple dashed line gives the luminosity function of \citet{Matsuoka2018} based on the $5.7<z<6.5$ quasar population and extrapolated to $z=7$.
    }
    \label{fig:UVLF}
\end{figure}
% fig ----------------------------------------------------------------------------------------------

\subsection{Dust extinction and UVLF of QSOs}

In this section, we evaluate the dust extinction of the QSO hosts in the UV band and make predictions for the corresponding UV luminosity function of QSOs.
We calculate the UV band dust optical depth $\tau_{\rm UV}$ for each line of sight surrounding the AGN and derive the dust-extincted UVLF taking into account the angular variations of $\tau_{\rm UV}$.
% Remember that we calculate the dust extinction based on the assumption that the dust extinction should be proportional to the metal column density along the line of sight through AGN, 
% and we use effective dust cross section $\sigma_{d} = 1.6\times 10^{-23} \rm{cm}^{-2}$ which is calibrated by the galaxy UVLF observation at z=7.

As illustrated in the middle column of Figure~\ref{fig:BH0map}, the dust optical depth $\tau_{\rm UV}$ has a similar pattern to the $N_{\rm H}$ map, indicating that a high level of dust extinction is more likely to happen for directions with high column density $N_{\rm H}$.
In Figure~\ref{fig:tau_NH} we give the 2D histogram of $N_{\rm H}$ of each line of sight surrounding the subset of the AGN population with $L_X > 10^{43}$ ergs/s at $z=7$, plotted against their $\tau_{\rm UV}$ values.
The calculation of $\tau_{\rm UV}$ is based on Eq.~\ref{equation:tau_dust} with the assumption that the dust extinction is proportional to the metal column density along the line of sight.
We see that for certain $N_{\rm H}$ values, (for example, $N_{\rm H} \sim 10^{23} \rm{cm}^{-2}$), the corresponding optical depths $\tau_{\rm UV}$  can sometimes be over 1.5 dex because of variations in the metallicity.
However, $N_{\rm H}$ and $\tau_{\rm UV}$ have an overall strongly positive correlation, indicating that the high density regions are more likely to yield higher levels of dust extinction. This in turn implys that the variations in metallicity along the lines of sight is subdominant compared to the density variations.

With knowledge of UV band extinction for each line of sight, we can predict the dust extincted UVLF of QSOs taking into account the angular variation of dust extinction.
The solid blue line in Figure~\ref{fig:UVLF} shows the intrinsic UVLF at $z=7$. 
To predict the dust extincted UVLF, we collect all the realizations of the dust extincted UVLF observed from the 972 different line of sights and take the average value, which we plot with a black solid line in Figure~\ref{fig:UVLF}. 
The grey shaded area gives the estimated error from the $2\sigma$ bounds of the 972 collections of dust extincted UVLFs. 

We can see that the dust extincted UVLF is about 1.5 dex lower than the intrinsic UVLF, implying that more than 99\% of the $z \sim 7$ AGNs are heavily dust extincted and might be missed by UV band observations.  
Observations of $z>7$ QSOs also lead to inference of a high level of obscuration in the UV band. 
Analysis based based on the UV spectra of two $z>7$ QSOs \citep{Davies2019} indicates that, given the total number of ionizing photons and the accreted black hole mass, the QSO could be obscured over more than 82\% of its lifetime (with the assumption of similar radiative efficiency as for low redshift QSOs). 
However, since our simulation does not keep enough snapshots to track the time evolution of all QSO hosts, we can not directly compare with this result by properly tracking the obscuration for specific QSOs on a time averaged basis.

In Figure~\ref{fig:UVLF}, we also show some observational data for high-$z$ QSO populations, using dashed and dotted lines.
The red dotted line shows the UVLF fitting result of \cite{Jiang2016}, which is based on observations of $z \sim 6$ QSOs. 
The orange data points with error bars are the measured binned UVLF from \cite{Wang2018}. 
The purple dashed line shows the QLF from \cite{Matsuoka2018} based on observations of $5.7<z<6.5$ QSOs, and the cyan dashed line is the estimate of \cite{Vito2018} inferred from x-ray observations.
Note that the X-ray band LF also includes the obscured AGN population while the UVLFs are only for optically selected QSOs.

As indicated by the blue and purple dashed lines, some observations of QSO luminosity functions favor a flattened feature at the faint end. To explore the possible factors that might affect the shape of the UVLF at faint end, we select the QSOs that have an intrinsic UV band luminosity higher than that of their host galaxy (i.e.,the points below the dashed line in Figure~\ref{fig:Muv_collection}) and plot the intrinsic and the corresponding dust extincted UV luminosity functions as green and red solid lines. 
This gives us an estimate of how the shape of UVLF changes if we take account of the fact that fainter AGN might be outshone by their host galaxy and therefore missed by UV observation.

% can outshine their host galaxy in UV band  
% To account for the fact that fainter AGN might be outshined by its host galaxy and therefore missed by UV observation, we also plot in green solid line the QLF that only incorporate the AGN population that can outshine their host galaxy in UV band (i.e.,the points below the dashed line in lower panel of Figure~\ref{fig:Muv_collection}). The red solid line corresponds to the dust extincted version of the green line.

%% file: Sec4_Conclusion.tex
\section{Summary}
\label{section4:Summary}

In this work, we study the galactic obscuration surrounding the $z=7$ QSO population based on the \textsc{BlueTides} cosmological hydrodynamic simulation, which in its latest run has just reached these redshifts.
We examine the angular variations of the gas obscuring AGNs as well as its correlation with the gas outflows driven by the AGN feedback. 
With the large size of the \textsc{BlueTides} simulation, we are able to study the obscuration state of QSOs on a statistical basis. 
We explore the relationship between the $N_{\rm H}$ distribution and AGN properties (intrinsic luminosity, BH mass and the Eddington accretion ratio) and host properties (gas outflow, stellar mass, molecular gas mass and the star formation rate), as well as tracing the time evolution of the overall $N_{\rm H}$ distribution. 
We also evaluate the UV band dust extinction and make predictions for the UV band QSO luminosity function.

Our main results are summarized below. Note that the statistics and linear regression fits below are calculated based on the QSO population with $L_X > 10^{43}$ ergs/s at $z=7$.

\begin{itemize}

\item For bright AGNs at $z>7$, galactic obscuration can exhibit large angular variations, spanning over 2 orders of magnitude for  different lines of sight. 
The angular directions with low column density $N_{\rm H}$ (i.e.,less obscured) are clearly correlated with the gas outflows driven by AGN feedback. 
Strong gas outflows can open up low column density regions with $N_{\rm H} \sim 10^{21} \mathrm{cm}^{-2}$.
Overall, for lines of sight with radial outflow velocities $v_r > 300$ km/s, 87\% have $N_{\rm H} < 10^{23} \rm cm^{-2}$ at $z=7$.
The morphology of the outflows typically show a bimodal structure, as indicated in Figure~\ref{fig:ang_prob}.

\item The host ISM for $z>7$ AGN is able to produce absorption up to Compton-thick level ($N_{\rm H} > 10^{24} \rm cm^{-2}$), which is in contrast with the case of low-$z$ AGN where Compton-thick obscuration can only be produced by parsec scale gas in the nuclear region.

\item For lines of sight with $N_{\rm H} > 10^{22} \rm{cm}^{-2}$, the obscuration is mostly contributed by high density clumps in the inner regions of galaxies, within $r<\rm{ckpc}/h$ of the central QSO. 
More quantitatively, on average we have $N_{\rm H} (<10 \rm{ckpc}/h)$ $\gtrsim 90\%$ $N_{\rm H} (<30 \rm{ckpc}/h)$.

\item The $N_{\rm H}$ distribution has a positive correlation with QSO luminosity, though when going to higher luminosity, $N_{\rm H}$ has a larger scatter and skews towards the low $N_{\rm H}$ end, driven by the AGN feedback. 
We fit a linear relationship between $\log N_{\rm H}$ and $\log L_X$, finding $\log N_{\rm H} = (0.42 \pm 0.02) \log L_{X} + (5.1 \pm 0.7)$.

\item The obscured fraction (defined as the fraction of sightlines to AGN with $N_{\rm H} > 10^{23} {\rm cm}^{-2}$) typically range from 0.7 to 1.0 for $L_X > 10^{43} \rm{ergs/s}$ AGN. 
% There is a slight trend that obscuration fraction is getting larger when AGN luminosity goes from $L_X \sim 10^{43}$ to $L_X \sim 10^{44}$ ergs/s. 
For AGN with higher luminosity ($L_X > 10^{44}$ ergs/s), the obscured fraction does not show a strong trend with the luminosity because  strong feedback driven by the bright QSOs  maintains the fraction of lines of sight with low $N_{\rm H}$. 

\item A similar trend is found to exist between the $N_{\rm H}$ distribution and BH mass, while the variations of $N_{\rm H}$ are larger for AGNs in the most massive $M_{\rm BH}$ bins. 
For BH masses with $M_{\rm BH} > 10^{7.7} M_{\odot}$, the probability of finding $N_{\rm H} < 10^{23} \rm{cm}^{-2}$ is 27\%. 
Linear regression between $N_{\rm H}$ and $M_{\rm BH}$ gives $\log N_{\rm H} = (0.13 \pm 0.02) \log M_{\rm BH} + (22.3 \pm 0.2)$.

\item The QSO population with large Eddington accretion ratio $\lambda_{\rm Edd}$ has an overall higher $N_{\rm H}$ distribution compared to the population with low $\lambda_{\rm Edd}$.
For QSOs with $\lambda_{\rm Edd} < 0.1$, the probability of finding $N_{\rm H} < 10^{23} \rm cm^{-2}$ is 40\%, while for $\lambda_{\rm Edd} > 0.7$, the probability is only 4\%.

\item There is no strong redshift evolution of the $N_{\rm H}$ distribution around QSOs above specific luminosity cuts.

\item With regard to host galaxy properties, the trend is for the AGN population in more massive system to have the overall $N_{\rm H}$ distribution peaking at higher values, but at at the same time becoming more broadened and skewed towards the low $N_{\rm H}$ end.
We split $N_{\rm H}$ by stellar mass, molecular mass and the star formation rate in the host, finding linear regression results of $\log N_{\rm H} = (0.24 \pm 0.03) \log M_{*} + (20.7 \pm 0.3)$, $\log N_{\rm H} = (0.47 \pm 0.03) \log M_{\rm H_2} + (18.4 \pm 0.3)$ and 
$\log N_{\rm H} = (0.48 \pm 0.02) \log (\rm sfr) + (22.31 \pm 0.04)$.

\item The dust optical depth $\tau_{\rm UV}$ has a tight positive correlation with $N_{\rm H}$. The regions with large dust extinction are more likely to have high $N_{\rm H}$, and the gas metallicity $Z$ only modulates the variation at a sub-dominant level.

\item Our dust extincted UVLF is about 1.5 dex lower than the intrinsic UVLF, implying that more than 99\% of the $z \sim 7$ AGNs are heavily dust extincted and will be missed by the UV band observations. 
We expect that up-coming and future X-ray missions (e.g., Athena \citep{Barcons2012}, Lynx \citep{Lynx2018}, AXIS \citep{AXIS2019}) will be able to reveal the hidden population (in UV band) of obscured AGN in the future. More detailed predictions for the number densities of QSOs expected for these deep X-ray surveys will be reserved to a follow-up study.

Finally we note that, given our findings that QSO driven feedback is the crucial factor for generating the low $N_{\rm H}$ tail (that allows the observation of the high-$z$ QSOs), it may be possible to use the inferred AGN obscured fraction at high-$z$ to provide some constraints for the strength of the AGN feedback. 
We expect the future observational samples of $z>7$ QSOs from the coming X-ray missions could provide a good test to the AGN feedback models.

% \yueying{The two main points brought by this work
% infers that the obscured fraction of high-$z$ AGN can provide a probe to the strength of the AGN feedback.
% -----------------------------------------------------------------------------------------
\end{itemize}